\RequirePackage{fix-cm}
\documentclass[twocolumn]{svjour3}          
\smartqed  
\usepackage{tabularx}
\usepackage{makeidx}  
\usepackage{amsmath,amssymb,amsfonts}
\usepackage[table]{xcolor}
\usepackage{scalerel}
\usepackage[graphicx]{realboxes}
\usepackage{xcolor}
\usepackage{colortbl}
\usepackage{xspace}
\usepackage[all]{nowidow}
\usepackage[inline]{enumitem}
\usepackage[scaled]{beramono}
\usepackage{booktabs} 
\usepackage[colorinlistoftodos]{todonotes}
\usepackage[tight,footnotesize]{subfigure}
\usepackage{multirow}
\usepackage{balance}
\usepackage{listings}
\usepackage{algorithm}
\usepackage{algpseudocode}
\usepackage{multirow}
\usepackage{tcolorbox}
\usepackage{color,hyperref}
\usepackage{pdflscape}
\usepackage{marvosym}
\usepackage{adjustbox}
\usepackage{tabularx}
\usepackage[figuresright]{rotating}
\usepackage{rotating}
\usepackage{csquotes}
\usepackage{xcolor}
\usepackage[flushleft]{threeparttable}

\lstdefinelanguage{m3}{
	basicstyle=\ttfamily\scriptsize,
	keywordstyle=\bfseries,
	keywords={m3,declarations,methodInvocation},
	literate={<-}{$\leftarrow$}{1},
	tabsize=2,
	alsoletter={-}
}

\newcommand{\code}[1]{{\small \texttt{#1}}}

\newcommand*\circled[1]{\tikz[baseline=(char.base)]{\color{black} 
		\node[shape=circle,draw=cyan,fill=black!10!white,inner sep=.3pt] (char) {{{\texttt\textbf #1}}};}}

\newtcolorbox{shadedbox}{
	drop shadow southeast,
	breakable,
	enhanced jigsaw,
	colback=white,
	boxrule=0.80pt
}

\usepackage{xspace}
\newcommand*{\ie}{i.e.,\@\xspace}
\newcommand*{\eg}{e.g.,\@\xspace}

\makeatletter
\newcommand*{\etc}{%
	\@ifnextchar{.}%
	{etc}%
	{etc.\@\xspace}%
}
\makeatother
\newcommand*{\etal}{\emph{et~al.}\@\xspace}

\newcommand\revised[1]{\textcolor{black}{#1}}
\newcommand\revision[1]{#1}

\newcommand*{\FaC}{FaCoY\@\xspace}

\newcommand*{\SO}{StackOverflow\@\xspace}
\newcommand*{\GH}{GitHub\@\xspace}

\newcommand*{\FC}{FOCUS\@\xspace}
\newcommand*{\AU}{AURORA\@\xspace}
\newcommand*{\MM}{MemoRec\@\xspace}
\newcommand*{\IES}{IE$_{s}$\@\xspace}
\newcommand*{\SES}{SE$_{s}$\@\xspace}
\newcommand*{\IEC}{IE$_{c}$\@\xspace}
\newcommand*{\SEC}{SE$_{c}$\@\xspace}

\newcommand{\rqfirst}{\textbf{RQ$_1$}: \revised{\emph{How well can \MM provide recommendations with different configurations?}\@\xspace}} 
\newcommand{\rqsecond}{\textbf{RQ$_2$}: \emph{How does the training data affect the performance of \MM?}\@\xspace}
	
\newcommand{\rqthird}{\textbf{RQ$_3$}: \emph{How do the encoding schemes affect the performance of \MM?}\@\xspace} 

\definecolor{darkgray}{gray}{0.78}
\definecolor{lightgray}{gray}{0.85}
\definecolor{verylightgray}{gray}{0.95}

\begin{document}


\title{A Context-aware Collaborative filtering Recommender System to Support the Specification of Metamodels} 

\title{MemoRec: A Recommender System to Support the Specification of Metamodels}

\title{MemoRec: A Recommender System for Assisting Modelers in Specifying Metamodels}



\author{Juri~Di~Rocco \and
 	    Davide~Di~Ruscio \and
 	    Claudio~Di~Sipio \and
	    Phuong~T.~Nguyen \and
       	Alfonso~Pierantonio
	}

\authorrunning{Di Rocco, Di Ruscio, Di Sipio, Nguyen and Pierantonio} 

\institute{	\Letter~Davide Di Ruscio \at
	Universit\`a degli studi dell'Aquila, Italy \\ 
	\email{davide.diruscio@univaq.it}
	\and
	Juri Di Rocco \at
	Universit\`a degli studi dell'Aquila, Italy \\ 
	\email{juri.dirocco@univaq.it}           
	\and
	Claudio Di Sipio \at
	Universit\`a degli studi dell'Aquila, Italy \\ 
	\email{claudio.disipio@graduate.univaq.it}	
	\and
	Phuong T. Nguyen \at
	Universit\`a degli studi dell'Aquila, Italy \\ 
	\email{phuong.nguyen@univaq.it}	
	\and
	Alfonso Pierantonio \at
	Universit\`a degli studi dell'Aquila, Italy \\ 
	\email{alfonso.pierantonio@univaq.it}	
}


\date{Received: date / Accepted: date}

\maketitle

\begin{abstract}
Model Driven Engineering (MDE) has been widely applied in software development, 
aiming to facilitate the coordination among various stakeholders. Such a 
methodology allows for a more efficient and effective development process. 
Nevertheless, modeling is a strenuous activity that requires proper knowledge 
of components, attributes, and logic to reach the level of abstraction required 
by the application domain. In particular, metamodels play an important role in 
several paradigms, 
and specifying wrong entities or attributes in metamodels can negatively impact 
on the quality of the produced artifacts as well as other elements of the whole process.
\revision{During the metamodeling phase, modelers can benefit from assistance to avoid mistakes, \eg getting recommendations like meta-classes and structural features relevant to the metamodel being defined. However, suitable machinery is needed to mine data from repositories of existing modeling artifacts and compute recommendations.} In this work, we propose \MM, a novel approach that makes use of a collaborative filtering strategy to recommend valuable entities related to the metamodel under construction. Our approach can provide suggestions related to both metaclasses and structured features that should be added in the metamodel under definition. We assess the quality of the work with respect to different metrics, \ie success rate, precision, and recall. The results demonstrate that \MM is capable of suggesting relevant items given a partial metamodel and supporting modelers in their task.
\keywords{Model-Driven Engineering \and Recommender Systems \and Collaborative Filtering Techniques}
\end{abstract}

\section{Introduction}
\label{sec:Introduction}
Model-Driven Engineering (MDE)~\cite{schmidt2006guest} development relies heavily on metamodels and models 
which are used to represent an abstraction of real-world entities as well as to produce application code automatically. As a result, the modeling activity represents the core of this para\-digm and should be carefully addressed to avoid possible errors in application deployment. Nowadays, modelers are equipped with 
several tools that support their tasks with different features, \ie graphical environment, drag-and-drop utilities, and auto-completion. 
So far, different modeling assistants have been proposed to support modelers in their daily activities 
\cite{batot_generic_2016,dupont_building_2018,lopez-fernandez_example-driven_2015,mora_segura_extremo_2019}. 
Nevertheless, most of them deal with testing or repairing~\cite{DBLP:conf/models/BarrigaRH18}, and to the best of our knowledge there have been no approaches dedicated to supporting metamodel specification. In particular, packages and metaclasses are the building blocks of a metamodel, but so far there exists no tool to help modelers effectively specify these artifacts. \revised{In fact, while working on a metamodel, modelers might expect to get recommendations consisting of relevant metaclasses or structural features that can be further integrated. However, due to a huge amount of available resources, searching for suitable artifacts is a daunting task. Under the circumstances, we see an urgent need for suitable machinery to mine data from open source platforms such as \GH. Among others, we are interested in finding which packages and metaclasses can be added to the metamodel under development, given that other packages and metaclasses are already defined.} 


%


\revised{In this work, we aim to provide modelers with an automated assistant, providing support during metamodeling activities. We propose \MM, a recommender system that exploits a context-aware collaborative filtering technique~\cite{Chen:2005:CCF:2154509.2154540} to recommend relevant artifacts related to the modeling domain. In particular, \MM has been conceptualized by learning from a series of recommender systems developed in the scope of the CROSSMINER project \cite{di_rocco_development_2021} to mine open source software, providing developers with various artifacts, including topics~\cite{10.1145/3382494.3410690}, API invocations, and source code~\cite{9359479}. \MM goes one step further to assist metamodeling activities by processing input data with four different encoding techniques (\ie different selections of what information has to be kept from a metamodel). More importantly, we tailor the internal design to compute similarity among metamodels in an efficient way.}  Given a metamodel partially specified by the modeler, \MM is able to suggest two types of artifacts, namely \emph{(i)} metaclasses at the level of package; and \emph{(ii)} structural features for a given metaclass. To properly capture the current context, we rely on the data encoding technique that has been successfully conceptualized in our previous work~\cite{8906979}. 
To our best knowledge, this is the first attempt to support MDE application development, exploiting collaborative filtering techniques. Thus, the contributions of our work are summarized as follows.
\begin{itemize}
\item \revised{A recommender system, named \MM, to provide modelers with classes and structural features relevant to the metamodels under development;}	
\item \revised{An empirical evaluation of the conceived system on two real metamodel datasets, employing a set of well-defined metrics commonly used in the recommender systems domain, \ie success rate, precision, recall, and F$_1$ score;}
\item The replication package of the tool has been made available to facilitate future research.\footnote{\url{https://github.com/MDEGroup/MemoRec}}
\end{itemize}


Through a careful observation, we realized that there \revised{exist no other tools that perform} the same tasks, and thus it is not possible to compare our tool with any baseline. In the scope of this paper, we evaluate the performance of our proposed tool by relying on the k-fold cross-validation technique~\cite{kohavi1995study}, aiming to investigate its practicality in real-world settings.


The paper is structured into the following sections. 
Section \ref{sec:Background}
shows a motivating example as well as \revised{background related to the context-aware collaborative filte\-ring technique.} In Section \ref{sec:ProposedApproach}, we present \MM and its main components. The evaluation is presented in Section \ref{sec:Evaluation}, while the results are analyzed in Section~\ref{sec:resutl}. \revised{A qualitative discussion of \MM is given in Section \ref{sec:discussion}. Section \ref{sec:RelatedWork} presents the work that is related to the approach presented in this paper.  We conclude and discuss possible future work in Section~\ref{sec:Conclusions}.}

\section{Motivations and Background}
\label{sec:Background}
\revised{This section describes the issues we tackle in this work and provides an  overview of the background. In particular, Section~\ref{sec:MotivatingExample} gives a motivating example where recommendations for completing a metamodel are needed. Afterward, Section~\ref{sec:focus} presents the context-aware colla\-borative-filtering technique, and its application in recommending API function calls and usage pattern as a base for further presentations.} 


\subsection{Motivating example}\label{sec:MotivatingExample}

\begin{figure*}[t!]
	\centering
	\begin{tabular}{c c }	
		\subfigure[The \textsc{Web} metamodel with the \code{Web} package and the \code{Page} class.]{\label{fig:mm_sample}
			\includegraphics[width=0.35\linewidth]{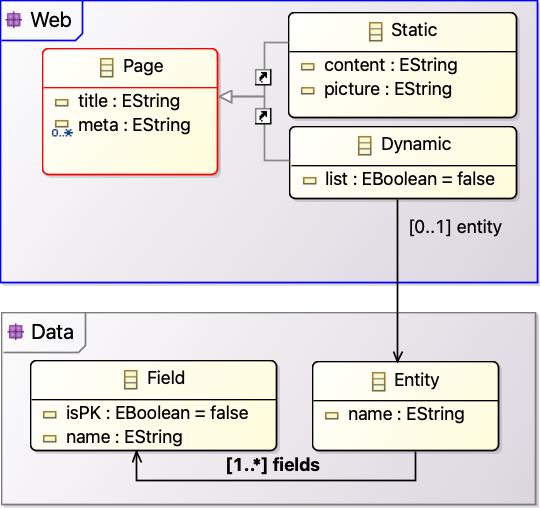}} &
		\subfigure[An example of recommendated elements for the active contexts.]{\label{fig:mm_sample_recs}
			\includegraphics[width=0.35\linewidth]{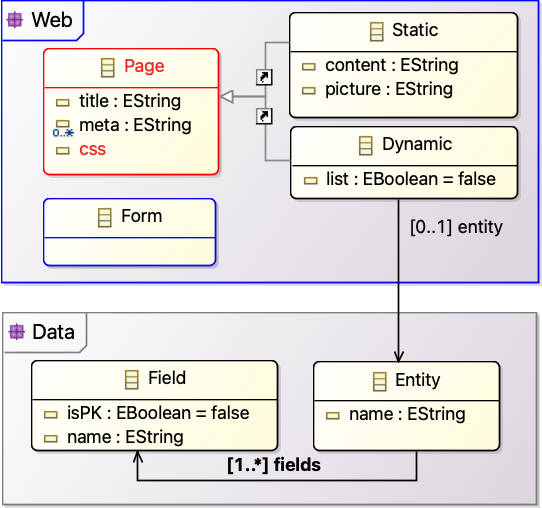}}	
	\end{tabular} 
	\vspace{-.2cm}
	\caption{\revised{The explanatory \textsc{Web} metamodel.}} 	
	\label{fig:motivating_example}
	\vspace{-.4cm}
\end{figure*}

\revised{We present a motivating example to illustrate the need for proper recommendations. The considered context is a modeler who is developing the Web metamodel depicted in 
Fig.~\ref{fig:mm_sample}. The metamodel consists of two packages, \ie~\code{Web} and \code{Data}, where the former models the presentation concepts and contains three classes, 
\ie~\code{Static} and \code{Dynamic} that inherit from \code{Page}. The latter groups data concepts such as~\code{Entity} and \code{Field}. By considering a metamodel and an active context as input, a model assistant is expected 
to recommend both additional classes and structural features that the metamodel under development should incorporate. An \emph{active context} is a package or a class that the modeler is attempting to define when asking for recommendations. In Fig.~\ref{fig:motivating_example}, let us suppose that the modeler is asking for additional classes and structural features for two given active contexts, \ie the \code{Web} package and the \code{Page} class, being marked with the blue and the red color, respectively.}


\revised{Fig.~\ref{fig:mm_sample_recs} depicts two possible recommendations for the given context, \ie~\code{css} structural feature (marked in red) within the \code{Page} class and \code{Form} class (marked in blue) within the \code{Web} package. To be concrete, 
the \code{Form} class allows the modeler to define complete HTML resources. By considering the possible structural feature recommendations  \code{Page} class, the usage of \code{css} adds common and interesting information to \code{Page}s in the Web domain. In other words, a modeler assistant should consider the active context to predict suitable recommendations that help the modeler complete the context.}

\subsection{\revised{Context-aware collaborative filtering technique}}\label{sec:focus}

\revised{A context-aware collaborative filtering   recommender system provides recommendations to users 
	items that have been bought by similar users in similar contexts~\cite{Sarwar:2001:ICF:371920.372071,Schafer:2007:CFR:1768197.1768208}. Based on this premise, we successfully developed a recommender system named FOCUS to provide developers with API function calls and usage patterns~\cite{Nguyen:2019:FRS:3339505.3339636}. We modeled the mutual relationships among projects using a tensor and mined API usage from the most similar projects.}

The successful deployment of different variants of collaborative filtering techniques allows us to transfer the acquired knowledge into the MDE domain. We build \MM by redesigning and customizing \FC to support the completion of metamodels. \revised{Mining from similar objects is a building block of collaborative filter\-ing techniques, and we exploit this to provide recommendations for metamodels under development. In particular, our approach works based on the assumption that: \emph{``if metamodels share some common artifacts, then they should probably have other common artifacts.''} In this respect, \MM mines meta-classes and structural features from similar metamodels, given an input metamodel. The following section presents our conceived approach to recommend useful elements for a metamodel being under development.}

\section{Proposed Approach}
\label{sec:ProposedApproach}



\revised{\MM ~provides ~modelers with~~recommendations, \newline which can be helpful while defining a metamodel, by considering the existing portion of the metamodel as active context. The system makes use of a graph representation to encode the relationships among various metamodels artifacts, and generates recommendations employing a context-aware collaborative filtering technique \cite{Schafer:2007:CFR:1768197.1768208}. In addition, we exploit a tailored textual representation of metamodels, which are encoded to enable the extraction of the containing knowledge to feed as input for the recommendation engine.} 

\begin{figure*}[t!]
	\centering
	\includegraphics[width=0.86\textwidth]{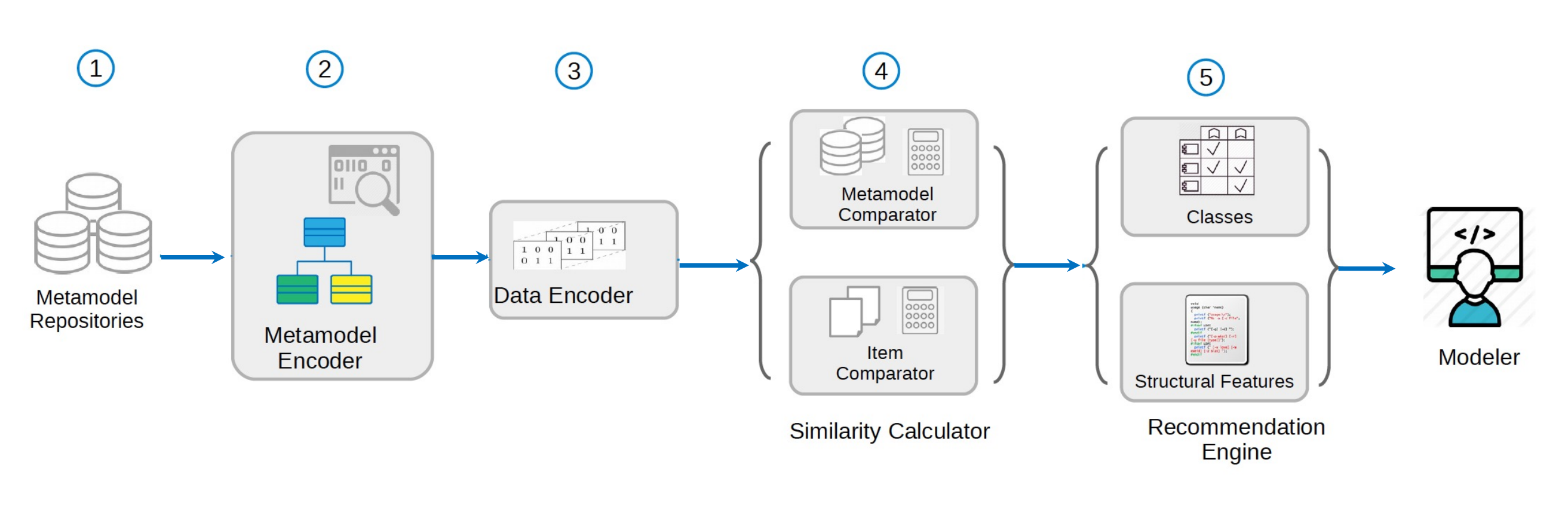}
	\vspace{-.2cm}
	\caption{Overview of \MM's architecture.} 
	\label{fig:Architecture}
	\vspace{-.2cm}
\end{figure*}


The architecture of \MM is depicted in Fig.~\ref{fig:Architecture}. To provide 
recommendations, \MM accepts as input a set of \textit{Metamodel 
Repositories}~\circled{1}. Afterwards, the \textit{Metamodel Encoder} component 
\circled{2} extracts packages and classes pairs as well as class and structural 
feature pairs from the metamodel being developed. \textit{Metamodel 
Comparator}, a subcomponent of \textit{Similarity Calculator} \circled{3}, 
measures the similarity between the metamodels stored in the repositories and 
the metamodel under specification. Using the set of metamodels and the 
information 
extracted by \emph{Metamodel Encoder}, the \textit{Data Encoder} component 
\circled{4} computes rating matrices. Given an active context of a metamodel, 
\ie package or class, \textit{Item Comparator} computes the similarities 
between packages and classes. From the similarity scores, 
\textit{Recommendation Engine} \circled{5} generates recommendations, either as 
a ranked list of classes if the active context is a packages, or a ranked list 
of structural features if the active context is a class. In the remainder of 
this section, we present in greater details each of these components.

\subsection{Metamodel Encoder}\label{sec:encodings}

A metamodel defines the abstract concepts of a domain where concepts, as well 
as the relationships among them, are expressed by the used modeling 
infrastructure. In particular, a metamodel consists of \textit{Packages} that 
aggregate similar concepts expressed by \textit{Classes}. \textit{Classes} 
consist of structural features, \ie attributes and references. Moreover, 
\textit{Classes} can inherit structural features from other classes.
%

\revision{Being inspired by our recent work~\cite{8906979}, we employ four encoding schemes to represent different views concerning the terms extracted from packages, classes, and structural features named instance. Each scheme has been used to elicit relevant information from the input metamodels according to different granularity levels.
 In particular, we make use of two \emph{standard encoding} (SE) scheme for recommending structural features within a context class and two other \emph{improved encoding} (IE) scheme for supporting classes within a package context. In particular, we consider the following definitions:}

\begin{itemize}
	\item \SES: it includes pairs in the form of \textit{$<$class\_nam\-e$>$\#$<$structural\_feature\_name$>$} for each structural feature contained within a class.  \revised{This encoding scheme is used to suggest additional structural features within a given class context};
	\item \IES: it consists of pairs \textit{$<$class\_name$>$\#$<$structur\-al\_feature\_name$>$} for each structural feature contained within a class. Moreover, it includes structural features  inherited from the super classes. \revised{In our previous work~\cite{10.1145/2593770.2593774}, we studied how structural features are used with hierarchies. On one hand, we found out that increasing the number of metaclasses with super-types decreases the average number of structural features directly specified in a metaclass, since structural features 
	are spread through the class hierarchies. On the other hand, the average number of structural features including the inherited ones is uncorrelated with the number of metaclasses with super-types. We anticipate that, by including the inherited structural features in \IES, we will be able to increase the informative part of the encoding. We use this scheme to suggest additional structural features within a given class context;}
	\item \SEC: it includes pairs \textit{$<$package\_name$>$\#$<$class\_na\-me$>$} for each class contained within a package.  \revised{This encoding scheme is utilized in providing additional classes within a given package context};
	\item \IEC: it flattens packages and classes and encodes classes within \revised{a default artificial package}. \revised{We use this encoding scheme to suggest additional classes within a given package context. Since metamodels consist of few \textit{ePackage}s~\cite{10.1145/2593770.2593774}, we envision that a flatten representation of classes, \ie by bypassing the Package/Class containment can help \MM to consider more metamodels and to extract 
	classes from different top similar metamodels.}
\end{itemize}

\noindent
\revised{An encoding scheme depends on two main factors:
\begin{itemize}
	\item[--] the purpose of the recommendation; depending on the type of the recommended items (\eg structural features, classes, specialization/generalization of a metaclass, \etc) the encoding scheme should be tailored to support the identified recommendation goal;
	\item[--] the prediction performance; the encoding scheme strongly impacts on the prediction performance. For this reason, the identification of suitable encoding scheme is an iterative process where the encodings are incrementally improved to maximize the prediction performance for a specific purpose.
\end{itemize}}

\revised{As future work, further encoding schemes could be provided to target different kinds of recommendations. For instance, an encoding scheme representing the inheritance relations between classes could suggest a possible set of generalizations or specializations for a given metaclass in the active context. In addition, a different encoding scheme could be used to include types for the recommended structural features.}

\begin{figure}[]
	\centering
	\begin{tabular}{c c }	
		\subfigure[SE$_{s}$ encoding]{\label{fig:SE_SF}
			\includegraphics[width=0.35\linewidth]{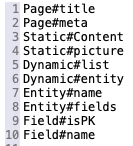}} 
		
		&
		\subfigure[IE$_{s}$ encoding]{\label{fig:IE_SF}
			\includegraphics[width=0.35\linewidth]{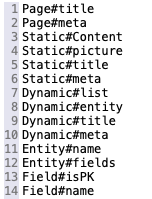}} \\

		\subfigure[SE$_{c}$ encoding]{\label{fig:SE_CLS}
			\includegraphics[width=0.35\linewidth]{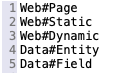}}		
		&
		\subfigure[IE$_{c}$ encoding]{\label{fig:IE_CLS}
			\includegraphics[width=0.35\linewidth]{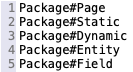}}
	\end{tabular} 
	\caption{\revised{The Web metamodel data extraction.}} 
	\label{fig:Encodings}
\end{figure}

Fig.~\ref{fig:Encodings} depicts an extract of the four encoding schemes 
related to the metamodels depicted in Fig.~\ref{fig:mm_sample}. In the 
following subsection, we show how the pairs \textit{package/class} and 
 \textit{class/structural feature} relationships are encoded.
\revised{In Section~\ref{sec:Evaluation}, we evaluated \MM by considering the four encoding schemes described in this section, \ie \SES, \IES, \SEC, and \IES.}

\subsection{Data Encoder}

Once package and class pairs as well as class and structural features have been extracted, \MM represents the relationships among them using two rating matrices to support class and structural feature recommendations. Given a metamodel, each row in the matrix corresponds to a package (class), and each column represents a class (structural feature). A cell is set to $1$ if the package (class) in the corresponding row contains the class (structural feature) in the column, otherwise it is set to $0$. 

Table~\ref{tab:pkg_cls_I} and Table~\ref{tab:cls_sf_I} illustrate how the metamodel depicted in Fig.~\ref{fig:mm_sample} is encoded into corresponding rating matrices. 
In particular, Table~\ref{tab:pkg_cls_I} shows the rating matrix combined with SE$_{c}$, whereas Table~\ref{tab:cls_sf_I} reports the rating matrix combined with IE$_{s}$.

\begin{table}
\caption{\revision{\emph{Package-class} feature rating matrix combined with SE$_{c}$ for the Web metamodel.}}
\begin{tabular}{|l|l|l|l|l|l|l|} \hline
	& Page & Static & Dynamic & Entity & Field \\ \hline
	Web & 1 & 1 & 1 & 0 & 0 \\ 
	Data & 0 & 0 & 0 & 1 & 1 \\ \hline
	
\end{tabular}

\label{tab:pkg_cls_I}
\end{table}

\begin{table}
	\caption{\revision{\emph{Class-structural feature} rating matrix combined with IE$_{s}$ for the Web metamodel.}}
	\begin{tabular}{|l|l|l|l|l|l|l|l|l|l|} \hline
		&\rotatebox{90}{title }& \rotatebox{90}{meta } & \rotatebox{90}{content } & \rotatebox{90}{picture } & \rotatebox{90}{list } & \rotatebox{90}{entity }& \rotatebox{90}{name }& \rotatebox{90}{fields } & \rotatebox{90}{isPK }\\ \hline
		Page & 1 & 1 & 0 & 0 & 0 &0&0&0&0\\ 
		Static & 1 & 1 & 1 & 1 & 0 & 0 & 0 & 0 &0\\ 
		Dynamic & 1 & 1 & 0 & 0 & 1 &1&0&0&0\\ 
		Entity & 0 & 0 & 0 & 0 & 0 & 0 &1 &1 & 0 \\ 
		Field & 0 & 0 & 0 & 0 & 0& 0& 1& 0& 1 \\ \hline		
	\end{tabular}
	\label{tab:cls_sf_I}
\end{table}



\begin{figure}[h!]
	\centering
	\includegraphics[width=0.45\textwidth]{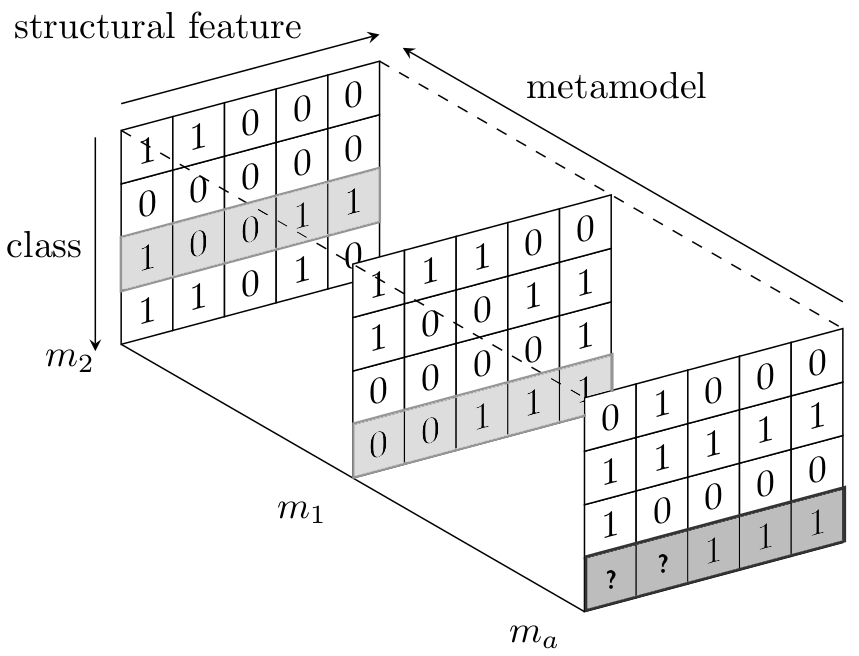}
	\caption{Matrix representation of metamodels w.r.t. structural features and classes.}
	\label{fig:3DRepresentation}
\end{figure}

A 3D context-based ratings matrix is introduced to model the intrinsic 
relationships among various metamodels, package (classes) and class (structural 
feature). The third dimension of this matrix represents a metamodel, which is 
analogous to the so-called context in context-aware collaborative filtering 
systems. For example, Fig.~\ref{fig:3DRepresentation} depicts three metamodels 
$M = (m_{a}, m_{1}, m_{2})$ represented by three slices with four classes and 
five structural features:
$m_{a}$ is the \emph{active metamodel} and it has an \emph{active context} \revision{highlighted in dark gray}. 
Both $m_{1}$ and $m_{2}$ are complete metamodels similar to $m_a$, and they are 
called \emph{background data}, as they serve as a base for the recommendation 
process. On one hand, the more background metamodel we have, the better is the 
chance that we recommend relevant structural features. On the other hand, 
increasing the number of top similar metamodels will enlarge the ratings 
matrix, and thus will add more computational complexity. 


\subsection{Similarity Calculator} \label{sec:SimilarityCalculator}
The recommendation of suitable metamodel items, \ie classes or structural features, is derived from similar metamodels and the active context, \ie packages or classes.
\textit{Similarity Calculator} is a generic component, it can be used to compute similarity for both classes and structural features. 
Given an active context of a metamodel under development, it is essential to find the subset of the most similar ones, and then the most similar contexts in that set of metamodels. 
\revised{Based on the active context type,} we create a weighted directed graph that models the relationships among metamodels and structural features to compute similarities.
\revised{Moreover, we implemented a graph-based similarity function~\cite{dirocco20,Nguyen:2019:FRS:3339505.3339636} to calculate the similarities among metamodels.}

\begin{figure}
	\centering
	\includegraphics[width=0.42\textwidth]{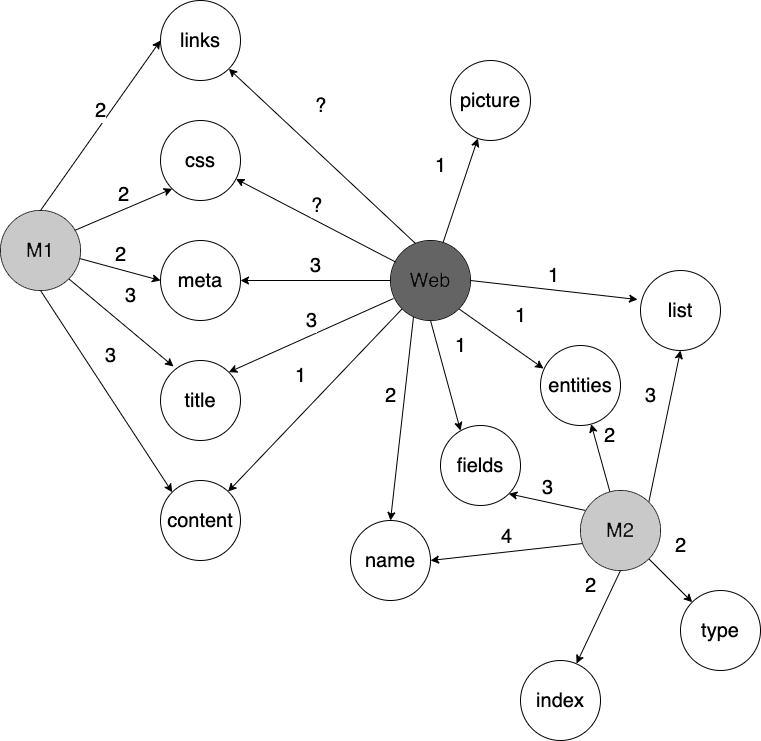}
	\caption{Graph representation of metamodels and structural features.}
	\label{fig:Similarity}
\end{figure}

In particular, we used two graph representations to support both class and 
structural feature recommendations. Each node in the graph represents either a 
metamodel or a structural feature. 
If metamodel $m$ contains structural feature $f$, then there is a directed edge from $m$ to $f$. The weight of the edge $m \rightarrow f$  corresponds to the number of times $m$ includes $f$. 
Figure~\ref{fig:Similarity} depicts the graph for the set of projects in 
Fig.~\ref{fig:3DRepresentation}: white nodes represent structural features, 
blue nodes represent most similar metamodels to the input ones depicted in 
green. 
For instance, the \textit{Web} metamodel has five classes and two of them 
define the attribute \textit{name}. As a result, the edge \textit{Web} 
$\rightarrow$ \textit{name} contains a weight of 2. In the graph, a question 
mark represents missing information, \ie for the active declaration in 
\textit{Web}, we need to find out if invocations \textit{links} and 
\textit{css} shall be included or not.


\revised{Given the node representing a metamodel $m$, there are nodes connected to $m$ via different edges, and they are called neighbor nodes.} Considering $(n_{1},n_{2},..,n_{l})$ as a set of neighbor nodes of $m$, the feature set of $m$ is the vector $\phi=(\phi_{1},\phi_{2},..,\phi_{l})$, where $\phi_{k}$ is the weight of node $n_{k}$, and computed using the \emph{term-frequency inverse document frequency} function \revised{computed by the following formula:
\begin{equation}
\phi_{k} = f_{n_{k}}*log(\frac{ \left | M \right |}{a_{n_{k}}})
\end{equation}}

\noindent
 where $f_{n_{k}}$ is the weight of the edge $m \rightarrow n_k$; \revised{$\left | M \right |$} is the number of all considered metamodels; and $a_{n_{k}}$ is the number of metamodels connected to $n_{k}$. 
 
The similarity between two metamodels $m$ and $n$ is comprehended as the cosine between their feature vectors $\phi=\{\phi_{k}\}_{k=1,..,l}$ and $\omega=\{\omega_{j}\}_{j=1,..,z}$, computed below:

\vspace{-.1cm}
\begin{equation} \label{eqn:VsmSim}
sim_1(m,n)=\frac{\sum_{t=1}^{\pi}\phi_{t}\times \omega_{t}}{\sqrt{\sum_{t=1}^{\pi}(\phi_{t})^{2} }\times \sqrt{\sum_{t=1}^{\pi}(\omega_{t})^{2}}} 
\end{equation}
\noindent where $\pi$ is the cardinality of the union of the sets of nodes by $m$ and $n$. 

Finally, the similarity between classes $c$ and $d$ is calculated with the Jaccard index given below: 
\begin{equation} \label{eqn:Jaccard}
sim_2(c,d)=\frac{|\mathbb{F}(c)\bigcap \mathbb{F}(d)|}{|\mathbb{F}(c)\bigcup \mathbb{F}(d)|} 
\end{equation}
where $\mathbb{F}(c)$ and $\mathbb{F}(d)$ are the sets of structural features for $c$ and $d$, respectively.

\revised{By referring to the motivation example proposed in Section~\ref{sec:MotivatingExample} and depicted in Fig.~\ref{fig:Similarity}, we present a concrete application of the proposed formalisms. Table~\ref{tab:phi-vector} reports the $\phi$ vectors for the \textit{Web}, \textit{M1}, and \textit{M2} metamodels. Then, Table~\ref{tab:similarity-matrix} lists the cosine similarity among vectors. Each cell reports the $sim_1$ score between the metamodels represented in the corresponding column and row. According to the results reported in Table~\ref{tab:similarity-matrix}, the \textit{Web} metamodel is more similar to \textit{M1} than to \textit{M2}}.

\begin{table*}
	\centering
	\caption{$\phi$ vectors for the metamodels depicted in Fig.~\ref{fig:Similarity}.}\label{tab:phi-vector}
	\begin{tabular}{|c|c|c|c|c|c|c|c|c|c|c|c|c|}
		\hline
		 & \textbf{links} & \textbf{css} & \textbf{media} & \textbf{title} & \textbf{content} & \textbf{picture} & \textbf{name} & \textbf{fields} & \textbf{entities} & \textbf{list} & \textbf{index} & \textbf{type} \\ \hline
		\textbf{Web} & 0 & 0 & 0.528 & 0.528 & 0 & 0 & 0 & -0.301 & -0.301 & -0.301 & 0 & 0 \\ \hline
		\textbf{M1} & 0 & 0 & 0 & 0.528 & 0.528 & 0 & 0 & 0 & 0 & 0 & 0 & 0 \\ \hline
		\textbf{M2} & 0 & 0 & 0 & 0 & 0 & 0 & 1.204 & 0.528 & 0 & 0.528 & 0.602 & 0.602 \\ \hline
	\end{tabular}
\end{table*}

\begin{table}
	\centering
	\caption{$sim_1$ matrix for the metamodels depicted in Fig.~\ref{fig:Similarity}.}\label{tab:similarity-matrix}
	\begin{tabular}{|c|c|c|c|}
		\hline
		& Web & M1 & M2 \\ \hline
		Web & 1 & \textbf{0.41} & -0.21\\ \hline
		M1 & \textbf{0.41} &1 & 0\\ \hline
		M2 & -0.21  & 0 & 1\\ \hline
	\end{tabular}
\end{table}

\subsection{Recommendation Engine}
This component is used to generate a ranked list of relevant items, \ie classes and structural features that depend on the metamodel context, \ie package and class.	
In the rest of this section, we present structural feature recommendations 
based on the class context. Analogously, we apply the same approach for recommending classes within packages.

Figure~\ref{fig:3DRepresentation} depicts an instance of structural features rating matrices.
In particular, the active metamodel $m_{a}$ already includes three classes, and the modeler is working on \revised{the fourth class}, corresponding to the last row of the matrix. The active class $c_{a}$ contains two structural features, represented in the last two columns of the matrix, \ie cells marked with $1$. The first two cells are filled with a question mark ($?$), implying that at the time of consideration, it is not clear whether these two structural features should also be added into $c_{a}$. \textit{Recommendation Engine} computes the missing ratings to predict additional structural features for the active class by exploiting the following collaborative filtering formula \cite{Chen:2005:CCF:2154509.2154540,Nguyen:2019:FRS:3339505.3339636}:
\vspace{1cm}
\revised{
\begin{equation} \label{eqn:missingRating}
r_{c,f,m} = \overline{r}_{c} + \frac{\sum_{d \in topsim(c)}(R_{d,f,m}-\overline{r}_{d}) \cdot sim_{2}(c,d)}{\sum_{d \in topsim(c)}sim_{2}(c,d)}
\end{equation}
}

Equation~\ref{eqn:missingRating} is used to compute a score for the cell representing structural feature $f$, class $c$ of metamodel $m$, where $topsim(c)$ is the set of top-N similar classes of $c$, $sim_{2}(c,d)$ is the similarity between two classes $c$ and $d$, computed by Equation~\ref{eqn:Jaccard}; 
$\overline{r}_{c}$ and $\overline{r}_{d}$ are calculated by averaging out all the ratings of $c$ and $d$, respectively;
$R_{d,f,m}$ is the combined rating of $d$ for $f$ in all the similar metamodels, computed in Equation~\ref{eqn:combinedRating}~\cite{Chen:2005:CCF:2154509.2154540}.
\revised{
\begin{equation} \label{eqn:combinedRating}
R_{d,f,m}=\frac{\sum_{n \in topsim(m)}r_{d,f,n} \cdot sim_{1}(m,n)}{\sum_{n \in topsim(m)}sim_{1}(m,n)} 
\end{equation}
}
\revised{\noindent where $topsim(m)$ is the set of top similar metamodels of $m$; and $sim_{1}(m,n)$ is the similarity between metamodels $m$ and $n$, calculated by means of Equation~\ref{eqn:VsmSim}.} Equation~\ref{eqn:combinedRating} suggests that given the active metamodel, a more similar metamodel is assigned a higher weight. This makes sense in practice, since similar metamodels contain more relevant structural features than less similar metamodels. Using Equation~\ref{eqn:missingRating} we compute all the missing ratings in the active class and get a ranked list of structural features with real scores in descending order. The list is then provided to modelers as recommendations.

\section{Evaluation}
\label{sec:Evaluation}
This section describes the datasets and the process we conceived to evaluate 
\MM. In particular, Section~\ref{sec:rqs} presents the research questions to study the performance of our proposed approach. Section~\ref{sec:dataset} gives an overview of the datasets used in the evaluation. The methodology and metrics are described in Section~\ref{sec:methodology} and Section~\ref{sec:Metrics}, respectively. 


\subsection{Research Questions}\label{sec:rqs}
The following research questions are considered to investigate \MM's recommendation performance: 

\begin{itemize}
	\item \rqfirst~We examine different configurations of \MM, \ie  the number of 
	recommended items as well as the number of neighbor metamodels, to find the 
	settings that bring the best performance. 
	\item \rqsecond~We study the outputs of \MM by considering two different datasets to assess to what extent their quality can have ripple effects on the prediction accuracy of \MM;	
	\item \rqthird~\revised{Since the definition of a suitable encoding scheme is an iterative process, we compared refined versions of the encodings, \ie \IEC and \IES, with the initial ones, \ie~\SEC~and \SES to pin down which of them facilitates the best recommendation outcome for \MM}.
\end{itemize}



\subsection{Data Extraction}\label{sec:dataset}

To evaluate the proposed approach, we exploited two independent datasets namely \textbf{D$_1$} and \textbf{D$_2$} as shown in Table~\ref{tab:dataset}, and they are described below.

	
\begin{itemize}
	\item \textbf{D$_1$} is a \textbf{curated} dataset~\cite{onder_babur_2019_2585456}, which consists of 555 metamodels mined from \GH and already labeled by humans. \revised{In particular, its metamodels have been already classified into nine categories, 
	\ie \emph{Bibliography}, \emph{Issue tracker}, \emph{Project build}, \emph{Review system}, \emph{Database}, \emph{Office tools}, \emph{Petrinet}, \emph{State machine}, and \emph{Requirements specification}. Though \MM does not require the input data to be labeled, such predefined categories are beneficial to the recommendation as there is a high similarity among the metamodels within a category. This is important since 
	MemoRec heavily relies on similarity to function (see Section~\ref{sec:ProposedApproach}), \ie given a metamodel, it searches for relevant items from similar metamodels;}

\item \textbf{D$_2$} is a \textbf{raw and randomly collected} dataset using the GitHub API \cite{githubAPI}. To aim for a reliable evaluation of \MM, \revised{we identified and filtered out from the dataset all the duplicated metamodels, resulting in a final set with 2,151 metamodels.}
\revised{By means of the \GH API \cite{githubAPI} we searched for files with the \code{.ecore} extension, which corresponds to Ecore metamodels. Due to the restrictions imposed by \GH, 
\eg it returns a maximum of 1,000 elements per query, we had to perform the searches by iteratively varying the query keywords. In particular, we used the \code{extension} qualifier as a base to search for \code{ecore} files. Then, we refined the query by adding typical \code{ecore} keywords, \ie~\code{ePackage}, \code{xml}, \code{eClass}, to name a few. Afterward, all the discovered metamodels were downloaded and collected in a dedicated folder. We removed all the files that we cannot directly parse with the EMF facilities~\cite{steinberg2008emf} from the corpus of collected artifacts. Finally, we removed the duplicated metamodels by the following process: \textit{(i)} a hash is computed for every collected \code{ecore} file based on its content; \textit{(ii)} the obtained hashes are used to build a hashmap where the key is the hash itself, and the value is the corresponding file; \textit{(iii)} if a duplicated key occurs in the map, we assume that the corresponding \code{ecore} is a duplicate and it is discarded.}

\end{itemize}






\begin{table}[t!]
	\caption{Datasets.}
	\begin{tabular}{|p{3.5cm}|p{1.5cm}|p{1.5cm}|} \hline
		 \textbf{Number of Artifacts} &  
		 \textbf{\textbf{D$_1$}} & 
		 \textbf{\textbf{D$_2$}}  \\ \hline
			Packages & 669  & 3,140 \\ 
		Metaclasses & 17,840  & 62,214 \\ 
		Structural features & 31,688 & 159,323 \\ 
			Attributes & 14,436  & 86,273 \\ 
				References & 17,252  & 73,050 \\ \hline
	\end{tabular}
\label{tab:dataset}
\vspace{-.4cm}
\end{table}
\subsection{Methodology}\label{sec:methodology}
As described in Section~\ref{sec:ProposedApproach}, \MM can recommend classes 
or structural features, \ie attributes and references, depending on the 
recommendation context, \ie packages or classes. For this reason, we perform 
the experiments by exploiting both classes and structural feature 
recommendations. In the rest of this section, we use classes within packages 
and structural features within classes as the recommendation objective.

\begin{figure*}[t!]
	\centering
	\includegraphics[width=0.90\linewidth,keepaspectratio]{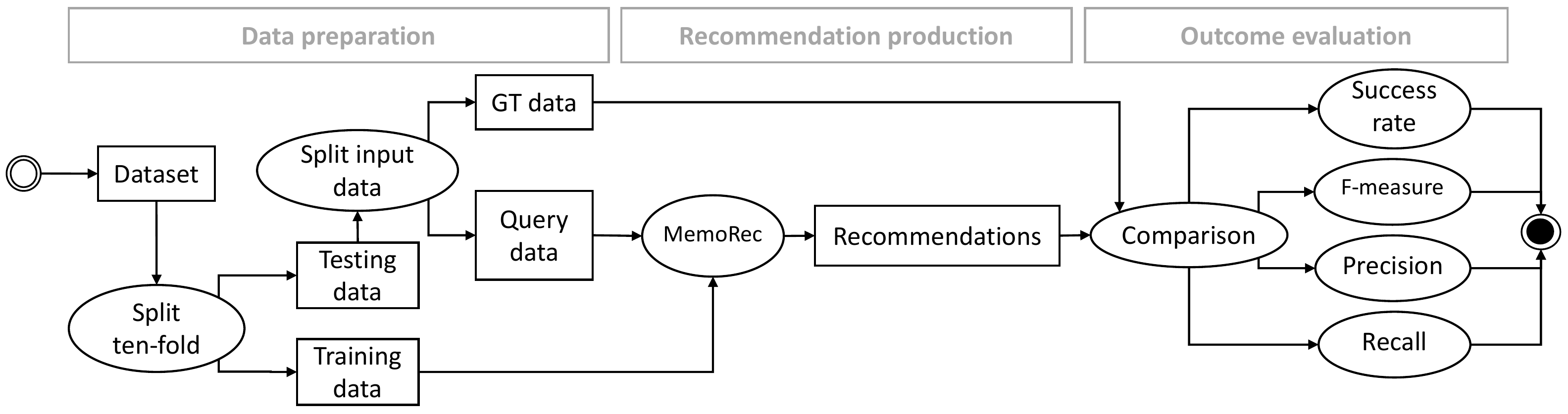}
	\caption{Evaluation Process.}
	\label{fig:EvaluationProcess}
\end{figure*}

To study if \MM is applicable in real-world settings, we perform an offline evaluation by simulating the behavior of a modeler who partially defines a metamodel and needs pratical recommendations on how to do next. Figure~\ref{fig:EvaluationProcess} depicts the evaluation process with three consecutive steps,~\ie~\emph{Data Preparation},~\emph{Recommendation Production}, and~\emph{Outcome Evaluation} explained as follows.

\begin{itemize}
	\item \textbf{Data Preparation.}
As seen in Fig.~\ref{fig:EvaluationProcess}, starting from an input \textit{Dataset}, we split it into two independent parts, \ie  \textit{Training data} and \textit{Testing data} (\textit{Split ten-fold}). The former corresponds to the metamodels collected ex-ante, whereas the latter represents the metamodel being modeled, or the active metamodel. 
The \textit{ten-fold} cross validation technique is used to conduct the evaluation as follows. The dataset is split into ten equal parts, one part represents the testing set and the remaining nine parts are combined to create a training set. 
We consider a modeler who is defining a metamodel \textit{m}, so some parts of \textit{m} are removed to mimic an actual metamodeling task: 
some packages/classes are already available in the active metamodel and the 
system should recommend additional packages/classes to be incorporated. 
\revised{For each metamodel in the \textit{Testing data}, by the \textit{Split input data} phase, a random package (class) that, together with the remaining
packages/classes, is selected to be used as \textit{Query data}. In the considered context, the first class/structural feature is kept as query data and all the others are taken out to be used as ground-truth (GT) \textit{data}.}
In other words, by the \textit{Split input data} phase, package (class) is selected as the active context \textit{c}. For \textit{c}, only the selected classes are provided as query, while the rest is removed and saved as ground-truth data. 



	\item \revised{\textbf{Recommendation production.} In this phase, the extracted \textit{Query data} and \textit{Training data} are fed as input for \MM, which in turn computes the final \textit{Recommendations}. It is important to remark that the current version of \MM can recommend classes and structural features. The types of the recommended attributes and relationships are not supported yet. This represents our next step to further develop the proposed approach.}
	\item \textbf{Outcome evaluation.} The performance of \MM is measured by comparing the recommendation outcomes with the ground-truth data (\textit{GT data}), exploiting the quality 
metrics, \ie \textit{Success rate}, \textit{Precision}, and \textit{Recall} which are presented in Section~\ref{sec:Metrics}. 
\end{itemize}

\subsection{Metrics}\label{sec:Metrics}
We introduce the following notations as a base for further presentation:
\begin{itemize}[noitemsep,topsep=0pt]
	\item \textit{N} is the cut-off value for the ranked list of recommendations. \revised{MemoRec returns a ranked list of recommendation items, and those on the top of the list are considered to be more relevant to the given context. The cut-off N is used to select the top items. For instance, if \MM retrieves four items, \eg~\newline \code{Form\-Element}, \code{Form}, \code{Link}, and \code{Page}, and the cut-off value $N$ is set to 2, then only the first two recommended items, \eg~\code{FormElement} and \code{Form} are provided as the final recommendation;}
	\item \textit{k} corresponds to the number of \textit{top-similar} neighbor metamodels \MM considers to predict suggested items;
	\item \textit{REC$_N$(m)} is the \textit{top-N} recommended items for \textit{m};
	\item \textit{GT(m)} is defined as the list of classes/structural features that are saved as ground-truth data for metamodel \textit{m};
	\item \textit{TP$_N$(m)} is the set of true positives, \ie items in the \textit{top-N} list that match with those in the ground-truth data, $TP_N (m) = GT (m) \bigcap REC_N (m)$.
\end{itemize}

Then the metrics, \ie success rate, precision, and recall are defined as follows.

\vspace{.1cm}
\noindent
$\rhd$~\textbf{SR@N.} Given the testing set of metamodels \textit{M}, success rate measures the ratio of queries that have at least a matched item among the total number of queries.
\begin{equation} \label{eqn:RecallRate}
SR@N=\frac{ count_{m \in M}( \left | TP_{N}(m) \right | > 0 ) }{\left | M \right |} 
\end{equation}
\revised{In particular \textit{SR} stands for success rate, and \textit{@N} corresponds to a cut-off value of N. For example, \textit{SR@2} means success rate for a recommended list with 2 items.}

\vspace{.1cm}
\noindent
$\rhd$~\textbf{Precision, Recall, and \revised{F$_1$-score (F-measure).}} 
These metrics are utilized to measure the \emph{accuracy} of the recommendation results. In particular, \emph{precision} corresponds to the ratio of \textit{TP$_N$(m)} among the number of recommended items,  \emph{recall} is the ratio of \textit{TP$_N$(m)} belonging to \textit{GT(m)}, \revised{whereas \emph{F$_1$-score} is the harmonic mean of precision and recall.}
\begin{equation} \label{eqn:Precision}
	\centering
precision = \frac{ \left | TP_{N}(m) \right | }{N}
\end{equation}
\vspace{-.2cm}
\begin{equation} \label{eqn:Recall}
recall = \frac{ \left | TP_{N}(m) \right | }{\left | GT(m) \right |}
\end{equation}
\revised{
\begin{equation} \label{eqn:f-measure}
	F_1 = 2*\frac{ precision * recall }{precision + recall}
\end{equation}}

\vspace{.1cm}
\noindent
$\rhd$~\textbf{Recommendation time.} To measure the time needed to perform a prediction, we used 
a laptop with 2,7 GHz Intel Core i7 quad-core 16GB RAM, and macOS Catalina 10.15.5.

\section{Experimental Results}
\label{sec:resutl}
This section analyzes the performance obtained by running \MM on the considered 
datasets. The three research questions are 
addressed in Section~\ref{sec:rq1}, 
Section~\ref{sec:rq2}, and 
Section~\ref{sec:rq3}. Section 
\ref{sec:threats} discusses possible threats 
to 
validity.


\subsection{\rqfirst}\label{sec:rq1}

We conducted experiments on the curated dataset (\ie \textbf{D$_1$}) by varying 
the number of neighbour nodes \textit{k} of the input metamodel, \ie $k = \{1, 
5, 10, 15, 20\}$, and the value of \textit{N}, \ie $N = \{1, 10, 20\}$. The 
rationale behind the selection of those values is as follows. First, we should 
not present a long list of recommended items since it may confuse the modeler, 
thus we select $N=20$ as the maximum value. Second, since the number of 
neighborhood items impacts on the computational complexity (cf. 
Equation~\eqref{eqn:combinedRating}), it is impractical to use a large 
number of metamodels as neighbors. Therefore, we consider the following values 
$k = \{1, 5, 10, 15, 20\}$. In the experiments, we use SE$_{c}$ and SE$_{s}$ as 
encoding schemes to compute recommendations for classes and structural 
features, respectively (the results related to the adoption of the other 
encoding schemes are presented in the discussion of RQ$_3$). 


Table~\ref{tab:structural_success_rate} and Table~\ref{tab:class_success_rate} 
show the average success rates obtained by running the ten-fold 
cross-validation technique to recommend structural features and classes, 
respectively by using different cut-off values $N$. In particular, 
Table~\ref{tab:structural_success_rate} depicts 
the success rate obtained for recommending structural features  classes. 
\revised{According to Table~\ref{tab:structural_success_rate}, we can observe that using more neighbour metamodels to compute recommendations brings a better success rate when the first recommendation item is considered.} For instance, \MM gets a success rate@1 of 0.153 and 0.202 
when k=1 and k=20, respectively. \revised{This is not confirmed when a longer list of recommendation items is considered, \ie $N = \{10, 20\}$.} Moreover, \MM yields a better performance when we 
increase the cut-off value N. Take as an example, for k=20, N=1, the obtained success rate is 0.202 which is less than a half of 0.479, the corresponding value when N=20. A longer list of recommended items means an increase in the match rate, however the modeler may tire of skimming through 
it. Thus, in practice, we should choose a suitable cut-off value N.


\begin{table}[h!]
	\caption{Success rate for structural feature recommendations, $k = \{1, 5, 
	10, 15, 20\}$, by considering the \textbf{D$_1$} dataset.}
	\centering
	\begin{tabular}{|l|c|c|c|}
		\hline
		\textbf{$k$} &  \textbf{SR@1}  &  \textbf{SR@10}  &  \textbf{SR@20}  \\ \hline 
		1                  & 0.152                      & 
		0.394                      & 0.501                   \\ 
		5                  & 0.181                     & 
		0.406                    & 0.488                      \\ 
		10                 & 0.202                     & 
		0.419                 & 0.502                     \\ 
		15                 & 0.200                 & 0.392                & 
		0.494                  \\ 
		20                 & 0.202                  & 0.362                & 
		0.479                    \\ \hline
	\end{tabular}
	\label{tab:structural_success_rate}
\end{table}

Table~\ref{tab:class_success_rate} report the success rate obtained with class 
recommendations. Partly similar to the results presented for recommending 
structural features  classes in Table~\ref{tab:structural_success_rate}, 
we see that incorporating more neighbors to compute recommendations is useful 
for a small k, \ie $k = \{1, 5, 10\}$. However, starting from k=15, there is a 
decrease in success rate by all the cut-off values N. We suppose that this 
happens due to the adoption of new neighbors that introduces only noise.

\begin{table}[t!]
	\caption{Success rate for class recommendations, $k = \{1, 5, 10, 15, 20\}$, by considering the \textbf{D$_1$} dataset.}%
	\centering
	\begin{tabular}{|l|c|c|c|} \hline
		\textbf{$k$} &  \textbf{SR@1}  &  \textbf{SR@10}  &  \textbf{SR@20}  \\ \hline
		1                  &    0.285          &   0.468       &  0.487                                   \\ 
		5                  &   0.204           &   0.617       &  0.690                                        \\ 
		10                 &   0.191           &   0.613       &  0.702                                   \\ 
		15                 &    0.215          &   0.583       &  0.694                             \\ 
		20                 &    0.187          &   0.539       &  0.661                  \\ \hline
	\end{tabular} 
	\label{tab:class_success_rate}
\end{table}



\medskip
\begin{tcolorbox}[boxrule=0.86pt,left=0.3em, right=0.3em,top=0.1em, bottom=0.05em]
	\small{\textbf{Answer to RQ$_1$.} \revised{Considering a certain number of similar metamodels contributes to 
			more relevant recommendations. Using data encoded with the \SEC and \SES encoding schemes allows \MM to predict better classes within a package than structural features within a class. Moreover, by considering a longer list of recommended items, \MM obtains an increase in success rate.}}
\end{tcolorbox}


\subsection{\revised{\rqsecond}}\label{sec:rq2}

We conducted similar experiments previously presented by measuring also the 
performance induced by the adoption of the dataset \textbf{D$_2$}. As 
previously described in Section \ref{sec:dataset}, 
\textbf{D$_1$} 
and \textbf{D$_2$} are different in terms of size and quality. In particular, 
\textbf{D$_1$} contains different groups of similar metamodels. Each group is 
labeled to refer the application domain that the metamodels in the considered 
group are intended to describe. Thus, as done for answering RQ$_1$, we 
performed experiments by varying the number of neighbour nodes of the input 
metamodel and the value of \textit{N}.

Table~\ref{tab:structural_success_rate_both_datasets} and 
Table~\ref{tab:class_success_rate_both_datasets} 
show the average success rates obtained by running the ten-fold 
cross-validation technique to recommend structural features and classes, 
respectively by using different cut-off values $N$.

\revised{According to Table~\ref{tab:structural_success_rate_both_datasets}, it is evident that using more neighbour metamodels to compute recommendations brings a better success rate for both datasets when the first recommended item is considered. An increasing number of neighbor $k$ does not improve the success rate values when a longer list of recommendations is considered, \ie $SR@10$ and $SR@20$.} 
However, by using the randomly created dataset \textbf{D$_2$} 
success rate is lower than that of \textbf{D$_1$}. For instance, with \textbf{D$_1$}, \MM gets a success rate@1 of 0.159 and 0.202 
when k=1 
and k=20, respectively, whereas with \textbf{D$_2$} the corresponding values 
are 0.114 and 0.161.  The same trend can also be seen with 
other cut-off values. As in the case of \textbf{D$_2$}, \MM yields a better 
performance when we increase the cut-off value N. Take as an example, with 
\textbf{D$_2$} and k=20, 
N=1, the corresponding success rate is 0.178 which is less than a half of 0.373, 
the corresponding value when N=20. In any case, the success rate related to the 
adoption of \textbf{D$_2$} is always lower than that of \textbf{D$_1$}.


\begin{table*}[t!]
	
	\centering

	\caption{Success rate, $k = \{1, 5, 10, 15, 20\}$, by comparing the adoption of \textbf{D$_1$} and 
		\textbf{D$_2$}}
	\label{tab:success_rate_rq2}
	
	\subtable[\footnotesize{for structural feature recommendations\label{tab:structural_success_rate_both_datasets}}]{
		\scriptsize
			\begin{tabular}{|l|l|l|l|l|l|l|}
			\hline
			\multirow{2}{*}{\textbf{$k$}} & \multicolumn{2}{c|}{\textbf{SR@1}} & 
			\multicolumn{2}{c|}{\textbf{SR@10}} & 
			\multicolumn{2}{c|}{\textbf{SR@20}} \\ \cline{2-7} 
			& \textbf{D$_1$}               & \textbf{D$_2$}               & 
			\textbf{D$_1$}                & \textbf{D$_2$}               & 
			\textbf{D$_1$}                & \textbf{D$_2$}               \\ \hline
			1                  & 0.153           & 0.114           & 
			0.394            & 0.328           & 0.502            & 
			0.397           \\ 
			5                  & 0.181           & 0.150           & 
			0.406            & 0.363           & 0.489            & 
			0.452           \\ 
			10                 & 0.202           & 0.161           & 
			0.419            & 0.340           & 0.501            & 
			0.413           \\ 
			15                 & 0.200           & 0.170           & 
			0.392            & 0.327           & 0.494            & 
			0.395           \\ 
			20                 & 0.202           & 0.178           & 
			0.362            & 0.320           & 0.479            & 
			0.373           \\ \hline
		\end{tabular}	
	}\qquad
	\subtable[\footnotesize{for class recommendations\label{tab:class_success_rate_both_datasets}}]{
		\scriptsize
		\begin{tabular}{|l|l|l|l|l|l|l|}
			\hline
			\multirow{2}{*}{\textbf{$k$}} & \multicolumn{2}{c|}{\textbf{SR@1}} & 
			\multicolumn{2}{c|}{\textbf{SR@10}} & 
			\multicolumn{2}{c|}{\textbf{SR@20}} \\ \cline{2-7} 
			& \textbf{D$_1$}               & \textbf{D$_2$}               & 
			\textbf{D$_1$}                & \textbf{D$_2$}               & 
			\textbf{D$_1$}                & \textbf{D$_2$}               \\ \hline
			1                  &    0.285      &    0.147      &    0.468    & 
			0.307       &   0.487               &   
			0.331                           \\ 
			5                  &   0.204        &   0.173    &    0.617   &   
			0.493      &    0.691               &    
			0.589                           \\ 
			10                 &   0.191       &    0.150     &    0.613   &   
			0.445     &   0.702               &        
			0.571                      \\ 
			15                 &    0.215     &     0.147    &    0.583    &  
			0.397      &    0.694                &              
			0.521               \\ 
			20                 &    0.187      &    0.141     &    0.539   &   
			0.362     &    0.661              &        0.478   \\ \hline
		\end{tabular}
	}

\end{table*}

Table~\ref{tab:class_success_rate_both_datasets} report the success rate 
obtained with class recommendations by comparing the adoption of  \textbf{D$_1$} and  
\textbf{D$_2$}. The decrease in accuracy related to the adoption of 
\textbf{D$_2$} as shown in Table~\ref{tab:structural_success_rate_both_datasets} is confirmed also in 
Table~\ref{tab:class_success_rate_both_datasets}.
\revised{To further study \MM's performance, we compute and report in 
Table.~\ref{tab:prec}, Table~\ref{tab:rec}, and Table~\ref{tab:fmeasure} the precision, recall, and f-measure \textbf{D$_1$} and \textbf{D$_2$}. 
For this setting, the number of recommended items \textit{N} was varied from 1 to 10, attempting to examine the performance for a considerably 
long list of items. 
The value of $k$ was varied of 1 to 20 with 5 as the step, considering a large number of 
neighbors.}



\begin{table*}[t!]

	\centering

	\caption{Precision values}
		\label{tab:prec}
	\vspace{-.3cm}
	\subtable[\footnotesize{ for structural feature recommendations using \textbf{D$_1$}\label{tab:PR_sf_d1}}]{
		\scriptsize
\begin{tabular}{|l|l|l|l|l|l|}
\hline
	\textbf{N} & \textbf{$K=1$} & \textbf{$K=5$} & \textbf{$K=10$} & \textbf{$K=15$} &\textbf{$K=20$} \\ \hline
1	 &	0.155 &		0.187 &		0.202	&	0.208 &	0.208 \\
2	 &	0.157 &		0.186 &		0.191	&	0.193 &	0.192 \\
3	 &	0.139 &		0.160 &		0.172	&	0.170 &	0.167 \\
4	 &	0.125 &		0.142 &		0.146	&	0.148 &	0.144 \\
5	 &	0.116 &		0.126 &		0.126	&	0.129 &	0.126 \\
6	 &	0.106 &		0.115 &		0.114	&	0.116 &	0.114 \\
7	 &	0.097 &		0.105 &		0.105	&	0.106 &	0.102 \\
8	 &	0.089 &		0.093 &		0.095	&	0.096 &	0.094 \\
9	 &	0.084 &		0.085 &		0.086	&	0.087 &	0.084 \\
10	 &	0.079 &		0.080 &		0.081	&	0.083 &	0.077 \\ \hline
\end{tabular}
	}\qquad
	\subtable[\footnotesize{for structural feature recommendations using \textbf{D$_2$}\label{tab:PR_sf_d2}}]{
		\scriptsize
\begin{tabular}{|l|l|l|l|l|l|}
	\hline
	\textbf{N} & \textbf{$K=1$} & \textbf{$K=5$} & \textbf{$K=10$} & \textbf{$K=15$} &\textbf{$K=20$} \\ \hline
1	 &	0.114 &		0.150 &		0.161	&	0.170 &	0.178 \\
2	 &	0.107 &		0.139 &		0.147	&	0.146 &	0.152 \\
3	 &	0.096 &		0.117 &		0.120	&	0.118 &	0.121 \\
4	 &	0.087 &		0.103 &		0.104	&	0.101 &	0.102 \\
5	 &	0.078 &		0.092 &		0.092	&	0.089 &	0.091 \\
6	 &	0.074 &		0.085 &		0.083	&	0.081 &	0.081 \\
7	 &	0.069 &		0.078 &		0.076	&	0.074 &	0.074 \\
8	 &	0.064 &		0.072 &		0.070	&	0.068 &	0.069 \\
9	 &	0.060 &		0.068 &		0.066	&	0.064 &	0.063 \\
10	 &	0.057 &		0.063 &		0.061	&	0.059 &	0.059 \\ \hline
\end{tabular}
	}
	\subtable[\footnotesize{for class recommendations using \textbf{D$_1$}\label{tab:PR_cls_d1}}]{
		\scriptsize
\begin{tabular}{|l|l|l|l|l|l|}
	\hline
	\textbf{N} & \textbf{$K=1$} & \textbf{$K=5$} & \textbf{$K=10$} & \textbf{$K=15$} &\textbf{$K=20$} \\ \hline
1	 &	0.236 &		0.193 &		0.204	&	0.198 &	0.180 \\
2	 &	0.235 &		0.213 &		0.179	&	0.182 &	0.164 \\
3	 &	0.218 &		0.198 &		0.162	&	0.159 &	0.142 \\
4	 &	0.209 &		0.181 &		0.156	&	0.149 &	0.138 \\
5	 &	0.196 &		0.175 &		0.151	&	0.144 &	0.124 \\
6	 &	0.185 &		0.168 &		0.150	&	0.134 &	0.114 \\
7	 &	0.177 &		0.160 &		0.147	&	0.131 &	0.112 \\
8	 &	0.166 &		0.148 &		0.142	&	0.130 &	0.114 \\
9	 &	0.163 &		0.151 &		0.144	&	0.127 &	0.110 \\
10	 &	0.156 &		0.158 &		0.146	&	0.125 &	0.105 \\ \hline
		\end{tabular}
	}\qquad
	\subtable[\footnotesize{for class recommendations using \textbf{D$_2$}\label{tab:PR_cls_d2}}]{
		\scriptsize
\begin{tabular}{|l|l|l|l|l|l|}
	\hline
	\textbf{N} & \textbf{$K=1$} & \textbf{$K=5$} & \textbf{$K=10$} & \textbf{$K=15$} &\textbf{$K=20$} \\ \hline
	1	 &	0.148 &		0.173 &		0.147	&	0.144 &	0.136 \\
	2	 &	0.138 &		0.165 &		0.152	&	0.142 &	0.132 \\
	3	 &	0.132 &		0.154 &		0.139	&	0.129 &	0.119 \\
	4	 &	0.123 &		0.144 &		0.127	&	0.118 &	0.107 \\
	5	 &	0.116 &		0.135 &		0.116	&	0.107 &	0.096 \\
	6	 &	0.111 &		0.128 &		0.110	&	0.101 &	0.090 \\
	7	 &	0.107 &		0.122 &		0.107	&	0.096 &	0.085 \\
	8	 &	0.103 &		0.117 &		0.102	&	0.093 &	0.081 \\
	9	 &	0.099 &		0.113 &		0.099	&	0.090 &	0.079 \\
	10	 &	0.096 &		0.112 &		0.096	&	0.087 &	0.076 \\ \hline

\end{tabular}
	}\quad

\end{table*}

\begin{table*}[t!]
	
	\centering

	\caption{Recall values}
	\label{tab:rec}
	\vspace{-.3cm}
	\subtable[\footnotesize{ for structural feature recommendations using \textbf{D$_1$}\label{tab:REC_sf_d1}}]{
		\scriptsize
\begin{tabular}{|l|l|l|l|l|l|}
	\hline
	\textbf{N} & \textbf{$K=1$} & \textbf{$K=5$} & \textbf{$K=10$} & \textbf{$K=15$} &\textbf{$K=20$} \\ \hline
1 & 0.054 &	0.063	&	0.070 &	0.070	& 0.071 \\
2 & 0.108 &	0.128	&	0.128 &	0.128	& 0.128 \\
3 & 0.142 &	0.161	&	0.174 &	0.168	& 0.163 \\
4 & 0.163 &	0.188	&	0.192 &	0.191	& 0.184 \\
5 & 0.182 &	0.203	&	0.202 &	0.204	& 0.195 \\
6 & 0.195 &	0.216	&	0.214 &	0.217	& 0.209 \\
7 & 0.205 &	0.227	&	0.225 &	0.225	& 0.214 \\
8 & 0.214 &	0.228	&	0.235 &	0.233	& 0.224 \\
9 & 0.228 &	0.233	&	0.239 &	0.239	& 0.226 \\
10 & 0.237 &	0.242	&	0.248 &	0.250	& 0.228 \\ \hline
\end{tabular}
	}\qquad
	\subtable[\footnotesize{for structural feature recommendations using \textbf{D$_2$}\label{tab:REC_sf_d2}}]{
		\scriptsize
\begin{tabular}{|l|l|l|l|l|l|}
	\hline
		\textbf{N} & \textbf{$K=1$} & \textbf{$K=5$} & \textbf{$K=10$} & \textbf{$K=15$} &\textbf{$K=20$} \\ \hline
 1 & 0.037	&	0.051	&	0.057	&	0.060	&	0.063 \\
 2 & 0.070	&	0.095	&	0.103	&	0.102	&	0.107 \\
 3 & 0.092	&	0.115	&	0.123	&	0.120	&	0.124 \\
 4 & 0.108	&	0.133	&	0.139	&	0.134	&	0.136 \\
 5 & 0.121	&	0.145	&	0.150	&	0.145	&	0.148 \\
 6 & 0.135	&	0.159	&	0.160	&	0.154	&	0.155 \\
 7 & 0.146	&	0.169	&	0.168	&	0.165	&	0.164 \\
 8 & 0.153	&	0.177	&	0.177	&	0.171	&	0.172 \\
 9 & 0.162	&	0.188	&	0.185	&	0.178	&	0.177 \\
 10 & 0.169	&	0.194	&	0.190	&	0.183	&	0.182 \\ \hline
\end{tabular}
	}
	\subtable[\footnotesize{for class recommendations using \textbf{D$_1$}\label{tab:REC_cls_d1}}]{
		\scriptsize
		\begin{tabular}{|l|l|l|l|l|l|}
			\hline
	\textbf{N} & \textbf{$K=1$} & \textbf{$K=5$} & \textbf{$K=10$} & \textbf{$K=15$} &\textbf{$K=20$} \\ \hline
1 & 0.032 & 0.039	&	0.048	&	0.049	&	0.048 \\
2 & 0.056 & 0.074	&	0.084	&	0.088	&	0.088 \\
3 & 0.067 & 0.091	&	0.102	&	0.111	&	0.109 \\
4 & 0.077 & 0.099	&	0.112	&	0.119	&	0.118 \\
5 & 0.083 & 0.109	&	0.120	&	0.132	&	0.127 \\
6 & 0.090 & 0.118	&	0.132	&	0.138	&	0.134 \\
7 & 0.099 & 0.124	&	0.143	&	0.149	&	0.144 \\
8 & 0.103 & 0.127	&	0.152	&	0.157	&	0.152 \\
9 & 0.109 & 0.137	&	0.162	&	0.163	&	0.158 \\
10 & 0.114 & 0.149	&	0.172	&	0.169	&	0.162 \\ \hline
			
		\end{tabular}
	}\qquad
	\subtable[\footnotesize{for class recommendations using \textbf{D$_2$}\label{tab:REC_cls_d2}}]{
		\scriptsize
		\begin{tabular}{|l|l|l|l|l|l|}
			\hline
	\textbf{N} & \textbf{$K=1$} & \textbf{$K=5$} & \textbf{$K=10$} & \textbf{$K=15$} &\textbf{$K=20$} \\ \hline
1	& 0.018	&	0.030	&	0.030	& 0.032	&	0.032	\\
2	& 0.034	&	0.058	&	0.062	& 0.063	&	0.061	\\
3	& 0.044	&	0.075	&	0.078	& 0.078	&	0.077	\\
4	& 0.050	&	0.085	&	0.087	& 0.085	&	0.083	\\
5	& 0.054	&	0.092	&	0.092	& 0.090	&	0.087	\\
6	& 0.058	&	0.099	&	0.098	& 0.096	&	0.092	\\
7	& 0.062	&	0.105	&	0.106	& 0.101	&	0.095	\\
8	& 0.065	&	0.110	&	0.111	& 0.106	&	0.099	\\
9	& 0.069	&	0.116	&	0.116	& 0.112	&	0.105	\\
10	& 0.072	&	0.123	&	0.121	& 0.116	&	0.109	\\ \hline
			
		\end{tabular}
	}\quad

\end{table*}

\begin{table*}[t!]
	
	\centering

	\caption{F-measure values}
	\label{tab:fmeasure}
	\vspace{-.3cm}
	\subtable[\footnotesize{ for structural feature recommendations using \textbf{D$_1$}\label{tab:fmeasure_sf_d1}}]{
		\scriptsize
		\begin{tabular}{|l|l|l|l|l|l|}
			\hline
			\textbf{N} & \textbf{$K=1$} & \textbf{$K=5$} & \textbf{$K=10$} & \textbf{$K=15$} &\textbf{$K=20$} \\ \hline
1&0.080&0.094&0.104&0.105&0.106 \\
2&0.128&0.152&0.153&0.154&0.154 \\
3&0.140&0.160&0.173&0.169&0.165 \\
4&0.141&0.162&0.166&0.167&0.162 \\
5&0.142&0.155&0.155&0.158&0.153 \\
6&0.137&0.150&0.149&0.151&0.148 \\
7&0.132&0.144&0.143&0.144&0.138 \\
8&0.126&0.132&0.135&0.136&0.132 \\
9&0.123&0.125&0.126&0.128&0.122 \\
10&0.119&0.120&0.122&0.125&0.115 \\\hline
		\end{tabular}
	}\qquad
	\subtable[\footnotesize{for structural feature recommendations using \textbf{D$_2$}\label{tab:fmeasure_sf_d2}}]{
		\scriptsize
		\begin{tabular}{|l|l|l|l|l|l|}
			\hline
			\textbf{N} & \textbf{$K=1$} & \textbf{$K=5$} & \textbf{$K=10$} & \textbf{$K=15$} &\textbf{$K=20$} \\ \hline
1&0.056&0.076&0.084&0.089&0.093 \\
2&0.085&0.113&0.121&0.120&0.126 \\
3&0.094&0.116&0.121&0.119&0.122 \\
4&0.096&0.116&0.119&0.115&0.117 \\
5&0.095&0.113&0.114&0.110&0.113 \\
6&0.096&0.111&0.109&0.106&0.106 \\
7&0.094&0.107&0.105&0.102&0.102 \\
8&0.090&0.102&0.100&0.097&0.098 \\
9&0.088&0.100&0.097&0.094&0.093 \\
10&0.085&0.095&0.092&0.089&0.089 \\ \hline
		\end{tabular}
	}
	\subtable[\footnotesize{for class recommendations using \textbf{D$_1$}\label{tab:fmeasure_cls_d1}}]{
		\scriptsize
		\begin{tabular}{|l|l|l|l|l|l|}
			\hline
			\textbf{N} & \textbf{$K=1$} & \textbf{$K=5$} & \textbf{$K=10$} & \textbf{$K=15$} &\textbf{$K=20$} \\ \hline
1&0.056&0.065&0.078&0.079&0.076 \\
2&0.090&0.110&0.114&0.119&0.115 \\
3&0.102&0.125&0.125&0.131&0.123 \\
4&0.113&0.128&0.130&0.132&0.127 \\
5&0.117&0.134&0.134&0.138&0.125 \\
6&0.121&0.139&0.140&0.136&0.123 \\
7&0.127&0.140&0.145&0.139&0.126 \\
8&0.127&0.137&0.147&0.142&0.130 \\
9&0.131&0.144&0.152&0.143&0.130 \\
10&0.132&0.153&0.158&0.144&0.127 \\ \hline
			
		\end{tabular}
	}\qquad
	\subtable[\footnotesize{for class recommendations using \textbf{D$_2$}\label{tab:fmeasure_cls_d2}}]{
		\scriptsize
		\begin{tabular}{|l|l|l|l|l|l|}
			\hline
			\textbf{N} & \textbf{$K=1$} & \textbf{$K=5$} & \textbf{$K=10$} & \textbf{$K=15$} &\textbf{$K=20$} \\ \hline
1&0.032&0.051&0.050&0.052&0.052 \\
2&0.055&0.086&0.088&0.087&0.083 \\
3&0.066&0.101&0.100&0.097&0.094 \\
4&0.071&0.107&0.103&0.099&0.093 \\
5&0.074&0.109&0.103&0.098&0.091 \\
6&0.076&0.112&0.104&0.098&0.091 \\
7&0.079&0.113&0.106&0.098&0.090 \\
8&0.080&0.113&0.106&0.099&0.089 \\
9&0.081&0.114&0.107&0.100&0.090 \\
10&0.082&0.117&0.107&0.099&0.090 \\ \hline
			
		\end{tabular}
	}\quad

\end{table*}


\revised{Table.~\ref{tab:prec} and Table.~\ref{tab:rec} confirm that by 
considering a curated dataset with more similar metamodels, \MM has better 
prediction performance than using a raw dataset. 
Moreover, \MM reaches better prediction performance in recommending structural 
features  classes than classes  packages.}

\begin{figure*}[h!]
	\centering
	\begin{tabular}{c c }	
		\subfigure[Dataset 
		\textbf{D$_1$}]{\label{fig:execution_time_D1}\hspace{-.3cm}\includegraphics[width=0.4\linewidth]{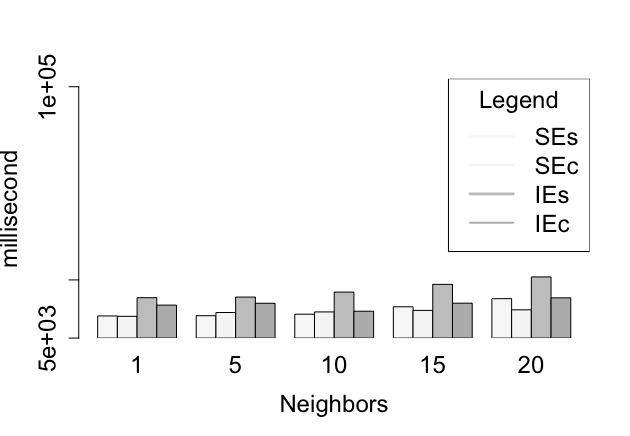}}
		 & 	
		\subfigure[	Dataset 
		\textbf{D$_2$}]{\label{fig:execution_time_D2}\hspace{-.3cm}\includegraphics[width=0.4\linewidth]{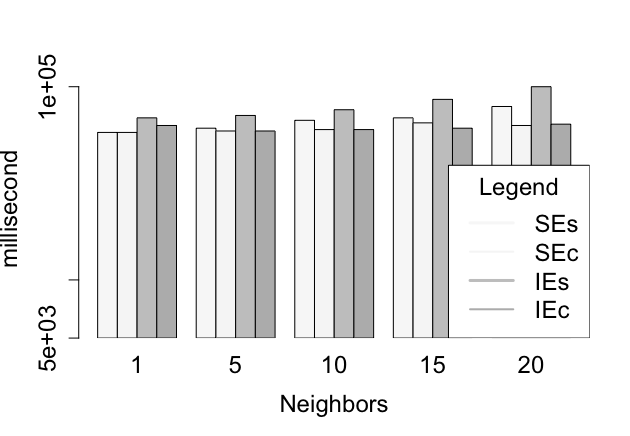}}
	\end{tabular} 
	\vspace{-.2cm}
	\caption{Average execution time.} 
	\label{fig:execution_time}
	\vspace{-.4cm}
\end{figure*}

\medskip
\begin{tcolorbox}[boxrule=0.86pt,left=0.3em, right=0.3em,top=0.1em, 
bottom=0.05em]
	\small{\textbf{Answer to RQ$_2$.} \revised{The quality of the input data plays a key role in \MM's performance. Curated datasets with more similar 
	metamodels allow \MM to improve its prediction performance, even if the size of such 
		datasets is smaller than that of those randomly collected.}} 
\end{tcolorbox}

\subsection{\revised{\rqthird}}\label{sec:rq3}

\begin{figure*}[t!]
	\centering
	\includegraphics[width=0.86\linewidth,keepaspectratio]{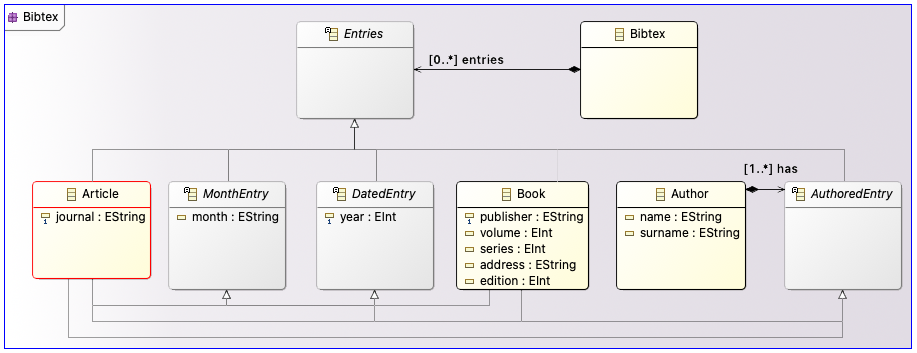}
	\caption{\revised{The \code{Bibtex} metamodel}.}
	\label{fig:evaluation_encoding}
\end{figure*}

\revised{Before addressing this research question, we discuss the recommendations produced by \MM on the example shown in Fig.~\ref{fig:evaluation_encoding}. More specifically, \MM has been applied on the \code{Bibtex} metamodel where the modeler is asking for the recommendation for two different contexts, \ie the \code{Article} class (rounded in red) and the \code{Bibtex} package (rounded in blue). Because of the different context's nature,  we show the recommendations to \code{Article} class by considering the \IES and \SES encodings. In contrast, the \IEC and \SEC encodings are discussed by considering the \code{Bibtex} package as the active context. We briefly summarize the four proposed schemes in Section~\ref{sec:encodings} as follows. \SES consists of \textit{class-structural feature} pairs for each structural feature directly defined within a class while \IES also includes structural feature inherited from the superclasses; \IEC consists of \textit{package-class} pairs while \SEC flattens packages within a default artificial package.} 


%
\revised{Table~\ref{tab:evaluation_encoding} shows the list of top-10 recommended items for each encoding scheme, \ie \SEC, \IEC, \SES, and \SEC. As described in Section~\ref{sec:encodings}, the type of the active context discriminates which encoding \MM needs to adopt.}	
\revised{By observing the outcomes of Table~\ref{tab:evaluation_encoding}, it can be seen that using \IES and \SES produces diverse recommended items. The result provided by \IES seems to contain more generic recommendations, \eg id, actor, attachment, to name a few, while \SES provides more focused recommended structural features. This intuition is confirmed by manually analyzing significant recommendation examples. Though both \IES and \SES provide useful suggestions to the modeler, 
the augmented information of \SEC allows \MM to predict more specific structural features. Referring to the scores obtained with \IEC and \SEC, we conclude that the results depend on the similarities among the training metamodels. By flattening package structure, \SEC mainly uses class names to compute similarities, while \IEC includes also package structure. So, when a similar metamodel is identified, \SEC can recommend classes that belong to any package.}

\begin{table}
	\centering
	\caption{Predicted recommendations for the \code{Article} metaclass and \code{Bibtex} package of the metamodel in Fig. \ref{fig:evaluation_encoding}.}
	\label{tab:evaluation_encoding}
	\begin{threeparttable}
		\begin{tabular}{|l||l||l|}
			\hline
			 & \textbf{\textbf{$IE_{s}$}  }         & \textbf{\textbf{$SE_{s}$}  }                           \\ \hline 
			{\multirow[origin=c]{10}{*}{\rotatebox[origin=c]{90}{\code{Article} metaclass} } } & containerPage  & address                                 \\ \cline{2-3} 
			& id                    & edition                                 \\ \cline{2-3} 
			& toolspecifics         & journal                                 \\ \cline{2-3} 
			& abstractPrefix        & month                                   \\ \cline{2-3} 
			& abstractText          & note                                    \\ \cline{2-3} 
			& acceptDirty           & number                                  \\ \cline{2-3} 
			& actor                 & pages                                   \\ \cline{2-3} 
			& address               & publisher                               \\ \cline{2-3} 
			& attachment            & series                                  \\ \cline{2-3} 
			& authorName            & volume                                  \\ \hline \hline
			
			 & \textbf{\textbf{$IE_{c}$}}           & \textbf{\textbf{$SE_{c}$} }                            \\ 
			\hline
			{\multirow[origin=c]{10}{*}{\rotatebox[origin=c]{90}{ \code{Bibtex} package } }} & AbstractCondition     & Abstract                                \\ \cline{2-3} 
			& Assign                & CommitteeMember
			\\ \cline{2-3} 
			& AssignExtra           & AcademiaOrganization                    \\ \cline{2-3} 
			& AttributeCondition    & AcademicEvent                           \\ \cline{2-3} 
			& Book                  & Academic\_Institution                   \\ \cline{2-3} 
			& ConditionalState      & AcceptRating                            \\ \cline{2-3} 
			& ConditionalTransition & Acceptance                              \\ \cline{2-3} 
			& Editor                & AcceptedPaper                           \\ \cline{2-3} 
			& EmphasisValue         & AccommodationPlace                      \\ \cline{2-3} 
			& Event                 & Account                                 \\ \hline
		\end{tabular}
	\end{threeparttable}
\end{table}

\vspace{.2cm}

\revised{To answer RQ$_3$, we conducted experiments on \textbf{D$_1$} by 
comparing pairwise the encoding schemes, \ie SE$_{s}$ versus
IE$_{s}$ and SE$_{c}$ versus IE$_{c}$.  As we mentioned before, we should select a suitable number of neighbor metamodels to compute recommendations to maintain a trade-off between efficiency and effectiveness. Thus, to simplify the evaluation, we set the number of neighbors \textit{k} to 5.\footnote{For each dataset and N value, the average success rate is higher with k$=5$ than that of other values of k.} The evaluation metrics computed by the encoding schemes for structural features and classes are reported in Table~\ref{tab:rq2_improvements} and Table~\ref{tab:rq2_improvements_class}, respectively.}

Table~\ref{tab:rq2_improvements} demonstrates an evident outcome: by using IE$_{s}$ as the encoding scheme, we obtain a better prediction performance than that when using SE$_{s}$. For instance, given $N=1$, we get a success rate of 0.241 and 0.181 for IE$_{s}$ and SE$_{s}$ encodings, respectively. Similarly by other evaluation metrics, \ie precision and recall, IE$_{s}$ helps achieve a superior performance. When we consider larger cut-off values, \ie $N = \{5, 10, 15, 20\}$, IE$_{s}$ is always beneficial to the recommendation outcomes, as it brings in higher quality indicators. Take as an example, with N=20, using IE$_{s}$ yields a success rate of 0.604, which is much higher than 0.489, the corresponding value for SE$_{s}$.


Next, we analyze the recommendations by using SE$_{c}$ and IE$_{c}$ as shown in Table~\ref{tab:rq2_improvements_class}. 
For success rate and precision, using IE$_{c}$ help \MM perform better than using SE$_{c}$. However, IE$_{c}$ negatively impacts on the recall values. In our opinion, it is due to the flattening operation which affects the similarity function, making metamodels too similar. In this case, the recommender engine is limited to suggests the most common classes. 
For instance, because of metamodelling best-practice, \code{NamedElement} is commonly used in a metamodel. This impacts on success rate and precision, but recall goes down because the recommended items do not depend on the metamodel context.

Through the experiment, we see that a suitable encoding scheme fosters better prediction performance. In this respect, we believe that the introduction of Natural Language Processing (NLP) steps, \ie stemming, lemmatization, and stop words removal, can boost up the accuracy of \MM. \revised{In particular, preprocessing steps can be employed to reduce the usage of different terms with very close semantics and, thus, to increase the corresponding term usages. For instance, the word ``\emph{reference}'' and its plural form ``\emph{references}'' are two conjugations of the same noun. Even though the two words have the same semantics, currently, \MM does not match those two terms and considers them different by negatively affecting the resulting \MM performance.}
\revised{We consider the integration of NLP techniques as our future work.}



\begin{table*}[t!]
	
	\centering
	\caption{Success rate, precision and recall for class recommendations}
	\label{tab:rq3}
	\subtable[\footnotesize{using SE$_{s}$ and IE$_{s}$ encodings.\label{tab:rq2_improvements}}]{
		\scriptsize
	\begin{tabular}{|l|l|l|l|l|l|l|}
	\hline
	\multirow{2}{*}{\textbf{N}} & \multicolumn{2}{c|}{\textbf{SR@N}}                                                          & \multicolumn{2}{c|}{\textbf{Precision@N}}                                                             & \multicolumn{2}{c|}{\textbf{Recall@N}}                                                                \\ \cline{2-7} 
	& \multicolumn{1}{c|}{\textbf{IE$_{s}$}} & \multicolumn{1}{c|}{\textbf{SE$_{s}$}} & \multicolumn{1}{c|}{\textbf{IE$_{s}$}} & \multicolumn{1}{c|}{\textbf{SE$_{s}$}} & \multicolumn{1}{c|}{\textbf{IE$_{s}$}} & \multicolumn{1}{c|}{\textbf{SE$_{s}$}} \\ \hline
	1                           & 24.074                                          & 18.113                                              & 0.241                                          & 0.181                                               & 0.078                                           & 0.061                                               \\ 
	5                           & 39.074                                          & 33.207                                              & 0.119                                           & 0.121                                               & 0.169                                           & 0.196                                               \\ 
	10                          & 48.333                                          & 40.566                                              & 0.084                                           & 0.078                                               & 0.241                                           & 0.245                                               \\ 
	15                          & 55.926                                          & 46.603                                              & 0.068                                           & 0.059                                               & 0.291                                           & 0.271                                               \\ 
	20                          & 60.370                                          & 48.867                                              & 0.058                                           & 0.047                                               & 0.331                                           & 0.284                                               \\ \hline
\end{tabular}
	}\qquad
	\subtable[\footnotesize{using SE$_{c}$ and IE$_{c}$ encodings.\label{tab:rq2_improvements_class}}]{
		\scriptsize
	\begin{tabular}{|l|l|l|l|l|l|l|}
	\hline
	\multirow{2}{*}{\textbf{N}} & \multicolumn{2}{c|}{\textbf{SR@N}}                                                          & \multicolumn{2}{c|}{\textbf{Precision@N}}                                                             & \multicolumn{2}{c|}{\textbf{Recall@N}}                                                                \\ \cline{2-7} 
	& \multicolumn{1}{c|}{\textbf{IE$_{c}$}} & \multicolumn{1}{c|}{\textbf{SE$_{c}$}} & \multicolumn{1}{c|}{\textbf{IE$_{c }$}} & \multicolumn{1}{c|}{\textbf{SE$_{c }$}} & \multicolumn{1}{c|}{\textbf{IE$_{c}$}} & \multicolumn{1}{c|}{\textbf{SE$_{c}$}} \\ \hline
	1                           & 22.593                                          & 20.370                                             & 0.226                                          & 0.204                                               & 0.014                                           & 0.037                                               \\ 
	5                           & 56.296                                          & 53.333                                              & 0.220                                           & 0.196                                               & 0.061                                           & 0.117                                               \\ 
	10                          & 65.926                                          & 61.667                                              & 0.194                                           & 0.163                                               & 0.100                                           & 0.152                                               \\ 
	15                          & 70.370                                          & 66.667                                              & 0.186                                           & 0.158                                               & 0.139                                           & 0.189                                               \\ 
	20                          & 73.333                                          & 69.074                                              & 0.167                                           & 0.140                                               & 0.159                                           & 0.210                                               \\ \hline
\end{tabular}
	}

\end{table*}


The average execution time among ten folds on various values of $k$ for both datasets is depicted in Fig.~\ref{fig:execution_time_D1} and Fig.~\ref{fig:execution_time_D2}. 
It can be seen that \revised{while IE$_{s}$ helps \MM achieve} a good prediction performance, it sustains a high computational complexity, resulting in prolonged execution time. This is understandable since compared to other techniques, the encoding scheme incorporates more information from metamodels for its computation.

\medskip
\begin{tcolorbox}[boxrule=0.86pt,left=0.3em, right=0.3em,top=0.1em, bottom=0.05em]
	\small{\textbf{Answer to RQ$_3$.} 
	\revised{
	Inherited structural features (IE$_{s}$) enable \MM to achieve a superior performance compared to using SE$_{s}$, despite a higher computational complexity. With IE$_{s}$, 
	\MM predicts better structural features within a class than classes within a package.}}
\end{tcolorbox}

\subsection{Threats to 
validity}\label{sec:threats}

In this section we give a discussion of  
threats, which might harm the validity of the 
performed experiments. In particular, we 
discuss threats with respect to internal and 
external validity as follows.

\noindent
\textbf{Internal validity.} Such threats 
refer to internal factors that might affect 
the outcomes of the performed experiments. A 
possible threat is represented by the 
datasets that have been used for the 
experiments. We mitigated such a threat by 
using two completely different datasets, and 
one of 
them has been randomly created without 
performing any data curation activities. 
Another internal threat to validity factor is 
represented by the adopted encoding schemes. 
Also in this case, we mitigated the issue by 
employing different encoding schemes. 
However, by considering data and encoding 
dimensions, we managed to identify 
distinctive characteristics of the approach 
that resulted to be valid independently from 
the adopted encoding schemes and input data sets, \ie graph builder, similarity calculator, and recommendation engine.

\noindent
\textbf{External validity.} It is related to  
factors that can affect the generalizability 
of our findings, by possibly making the 
obtained results not valid outside the scope 
of this study. We mitigated the issue by 
evaluating \MM in different 
scenarios, with the aim of simulating several 
usages of the systems, \eg by varying the 
number of neighbour metamodels,  and the size 
of the list of recommended items. \revised{Another threat to validity can be the fact that currently, we do not consider the sequences of actions that are operated to lead to a given metamodel. We believe that alternative approaches like LSTM (Long Short-Term Memory) \cite{10.1162/neco.1997.9.8.1735} can be a possible candidate to produce recommendations that rely on creation sequences.} \revised{Moreover, it is crucial to investigate how modelers perceive \MM. In this respect, we plan to conduct a user study where human evaluators are asked to give their assessment for recommendations provided by \MM.}

\section{\revised{Discussion}}
\label{sec:discussion}

Mussbacher \etal~\cite{10.1145/3417990.3421396} proposed a conceptual framework to characterize and compare approaches for intelligent modeling assistance (IMA). The proposed assessment grid consists of nine properties, namely \emph{Model}, \emph{Autonomy}, \emph{Relevance}, \emph{Confidence}, \emph{Trust}, \emph{Explainability}, \emph{Quality Degree}, \emph{Timeliness}, and \emph{Quality regarding external sources}. 
We provide a qualitative discussion of \MM with respect to such properties as follows.  

With \emph{Model}, the authors~\cite{10.1145/3417990.3421396} refer to the quality level of the models recommended by the considered IMA. Such a quality is measured with syntactic, semantic, and pragmatic quality. Concerning such a dimension, \MM ensures syntactic quality as long as it is trained with syntactically correct artifacts.  

With the \emph{Autonomy} property, authors refer to what extent the IMA under analysis is autonomous in gathering context information or user feedback without any user intervention. Such a property is mainly related to the tool gearing the considered IMA. \MM cannot be analyzed concerning the \emph{Autonomy} dimension because, at this stage, we focused on the algorithmic part of the approach and deferred its integration in a supporting environment as a next step.

The \emph{Relevance} property is related to the degrees of precision and recall of the recommendations provided by the adopted IMA with respect to modeler intensions. As shown in Section \ref{sec:resutl}, \MM has been evaluated by resembling different configurations, and the obtained accuracy is satisfactory, and it very much depends on the quality of the training data.

With the \emph{Confidence} property, 
Mussbacher \etal~\cite{10.1145/3417990.3421396} aim at measuring how often the IMA under analysis provides a confidence value for the recommended items. Similarly to all the analyzed IMAs~\cite{10.1145/3417990.3421396}, \MM does not provide any confidence value for the recommended metamodel elements. Indeed, this represents an interesting and important improvement for \MM.

\emph{Trust} has been defined as \emph{``the perception that the modeler has about the quality of an IMA.''} This represents an important property that we plan to assess once we have to evaluate the quality of the supporting tool integration of \MM in an existing IDE, \eg Eclipse.

The \emph{Explainability} property is another essential characteristic that is related to the emerging \emph{Explainable AI} research field \cite{DBLP:conf/icaart/Giannotti21}, which aims at existing AI techniques and tools to make produced outcomes understood by humans. The current version of \MM cannot provide any explanations related to recommended elements. An interesting extension can be complementing recommended metamodel elements with the sources of the most representative metamodels that triggered the given recommendations.

The \emph{Quality Degree} dimension measures \emph{``the degree of excellence of the IMA to address the needs of a modeler.''} Similar to all the approaches that have been analyzed \cite{10.1145/3417990.3421396}, \MM relies on external sources to provide modelers with recommendations that are valid for the active context.

Since \emph{Timeliness} refers to the user satisfaction for a given IMA, such a quality metric cannot be assessed for \MM at this stage. Instead, we plan to evaluate it once \MM is integrated into an existing IDE, like Eclipse.

The \emph{quality regarding external sources} property refers to the quality of the IMA under analysis concerning its external sources. As shown in Fig. \ref{fig:Architecture}, \MM is repository independent as long as it is possible to download the available modeling artifacts. However, as also discussed in Section \ref{sec:rq2}, being \MM a data-driven approach, the quality of the mined metamodels has an impact on the performance of \MM. Consequently, if the system is trained with low-quality metamodels, also recommendations would be of limited quality. 
\color{black}

\section{Related Work}
\label{sec:RelatedWork}


\subsection{\revised{Existing modeling assistants} \label{sec:ModelingAssistants}}


\revised{The Extremo tool~\cite{mora_segura_extremo_2019} has been proposed to assist modelers in a platform-independent way. By relying on miscellaneous resources excerpted from the context, it creates shared data employed to develop a flexible query mechanism. This querying system is used to explore and find useful entities that help the modeler complete the model under construction.  To assess the quality of the work, Extremo has been used to implement a DSL for the financial domain and validate its soundness through different use cases. The tool has been fully integrated into the Eclipse IDE as a plugin. In contrast with \MM that extracts the query directly from the metamodel under development, Extremo requires a modeler to perform custom or pre-defined queries to search for information chunks.}

\revised{Papyrus~\cite{dupont_building_2018} is a system to support  
domain-specific model specification by exploiting UML profiles. The mapping between each profile's metaclass and real Java classes is performed using EMF generator model utilities. Furthermore, the tool embeds a palette and a context menu to graphically specify the selected metaclasses. Though the results are promising, more features could be added to enrich the experience, \eg proactive triggering of recommendations or fine-grained customizations using EMF utilities. By transforming UML profiles to EMF metamodels, Papyrus supports the development of domain-specific environments. Although the generated domain-specific environments facilitate editing a model, Papyrus does not support the modeler with suggestions to complete the input profile from which the EMF metamodel is generated nor edit a model that conforms to those specifications.}

\revised{Batot and Sahraoui \cite{batot_generic_2016} 
introduced a modeling assistant based on a 
multi-objective optimization problem (MOOP). 
A well-founded evolutionary algorithm, namely 
NSGA-II, is employed to obtain representative 
models using an initial set provided by the 
user. According to the Pareto optimality 
definition, the algorithm is able to solve 
the MOOP to find relevant candidates. In this 
way, NSGA-II retrieves partial models to be 
completed by expert-domain users who can 
personalize the obtained results by selecting 
different coverage degree or changing 
pre-defined minimality criteria. To assess 
the quality of the work, the proposed NSGA-II 
adaptation was compared with random and 
mono-objective functions. 
Experimental results showed that the MOOP adaptation outperforms the baselines. 
Different from \MM that suggests possible metamodel elements, the approach in \cite{batot_generic_2016}  generates a set of models for various MDE tasks, \eg, testing automated learning.}

\revised{López-Fernández \etal~\cite{lopez-fernandez_example-driven_2015} presented an example-driven tool to 
recommend a complete metamodel starting from model fragments specified by graphical tools, \eg Visio, PowerPoint, Dia. The tool extracts untyped model fragments using initial examples as the starting point. Then, it infers an agnostic metamodel from them that the modeler can then enrich. 
In contrast with \MM, metamodels are generated by an iterative and inductive process where model fragments are given either sketched by domain experts using drawing tools or by a compact textual notation.}

\revised{AVIDA-MDE \cite{goldsby_avida-mde_2008} extends the original AVIDA tool and facilitates the generation of behavioral models starting from the requirements' specification. 
At the beginning of the process, an instinctual knowledge is formed by considering state diagrams, their inner elements, and the alphabet to specify such models. By relying on such information, AVIDA-MDE constructs new transitions and propose different behavior models to support scenarios and meet the constraints specified in the initial phase by the user. The approach was evaluated using a robot navigation system as the testing scenario. AVIDA-MDE aims at generating a set of behavioral models starting from the given parameters, whereas \MM recommends additional model elements starting from the context consisting of the metamodel under development.} 

\revised{A model assistant based on clone detection has been recently envisioned \cite{stephan2019towards}. The envisioned prototype is able to edit incomplete input model as well as propose fine-grain operations on the model itself. The approach makes use of Simulink Virtual Modeling Assistant (SimVMA), a well-founded technique to detect model clones. SimVMA is capable of finding all possible intersections between the initial metamodel and the clones belonging to the knowledge base. By exploiting the Type-3 clone similarity, the system can perform the two recommended types, \ie, retrieving complete models or single operation suggestions. By using similar metamodels, \MM applies a context-aware collaborative filtering technique using a given number of similar metamodels to predict additional model elements.}

\revised{MAR~\cite{10.1145/3365438.3410947} employs a query-by-example approach to search for similar metamodels/models. First, model structure is encoded as bags of paths before being indexed and stored on Apache HBase. Given a model, MAR uses it as a query to search for similar artifacts using a similarity score. Though modelers can learn by inspecting similar metamodels, they have to manually examine the results to extract useful information.}

\revised{Sen \etal \cite{doi:10.1177/0037549709340530} proposed a model assistant based on constraint logic program (CLP) to support the definition of domain-specific models in the modeling framework Atom$^3$. Given a partial model, the proposed tool is able to synthesize the complete model by relying on several constraints specified in Prolog. The designer can eventually ask for additional recommendations using the generated domain-specific model editor. Similarly, an extension of Diagram Predicate Framework (DPF) has been proposed to rewrite partial models by adding new elements graphically \cite{Rabbi2015ADA}. In particular, the proposed framework grants the compliance of the edited model using several termination rules adapted from the layered graph grammar technique. Differently from \MM that mines existing metamodels to predict additional model elements, the proposed approach does not use information from existing models, but makes use of \emph{i)} the notation of the modeler under development, \emph{ii)} the constraints expressed on this metamodel, and \emph{iii)} the partial model built by a domain expert to generate a visual model editor for the DSML supporting recommendations for possible completions.} 

\revised{The ASketch tool \cite{Wang2018ASketchAS} supports the completion of Alloy partial model with holes using automated analysis. First, the input interpreter parses the partial model to generate possible candidates. Afterward, these model fragments are encoded with the partial model and AUnit test files. Thus, ASketch can find possible solutions in a large search space by relying on an SAT solver. ASketch fills an Alloy partial model with concrete candidate fragments such that predefined tests, \ie unit testing, test execution.
Therefore, the final model fragments are not extracted from existing model/metamodel, which is actually done by \MM.}	

\revised{Recently, we presented MORGAN~\cite{morgan21}, a recommender system based on a graph neural network (GNN) to assist modelers in performing the specification of metamodels and models.
Similar to \MM, MORGAN makes use of tailored model and metamodel parsers to excerpt relevant information in textual data format. Then, the encoder builds the graphs from the text produced by the parser. Finally, the generated graphs feed a GNN-based engine to compute additional metamodel or model parts. Unlike \MM, MORGAN does not need an active context where the recommender suggests additional elements. At the same time, \MM attempts to get very related recommendations to the context where the modeler is working on the definition. For this reason, we cannot directly compare MORGAN with MemoRec.}

\revised{Our work distinguishes itself from the studies mentioned above as it can provide missing classes and structural features for a metamodel under development, exploiting a context-aware collaborative filtering technique. In addition, we anticipate that applying natural language processing techniques can help \MM improve the prediction performance, and we consider the issue in our future work.}


\subsection{\revised{Code recommender systems}} \label{sec:RecSys}

	
\revised{In the context of open-source software, developing new systems by reusing existing components raises relevant challenges in \textit{(i)} searching for relevant modules;  and \textit{(ii)} adapting the selected components to meet pre-defined requirements. To this end, recommender systems in software engineering have been developed to support developers in their daily tasks~\cite{di_rocco_development_2021,robillard_recommendation_2014}.  Such systems have gained traction in recent years as they can provide developers with a wide range of valuable items, including code snippets~\cite{Nguyen:2019:FRS:3339505.3339636,9359479}, tags/topics \cite{10.1145/3383219.3383227}, third-party libraries \cite{NGUYEN2019110460}, documentation~\cite{RUBEI2020106367}, to mention but a few.} 

Sourcerer \cite{linstead_sourcerer:_2009} performs code search in large-scale repositories by exploiting different components. The fir\-st component is the crawler, which automatically downloads repositories to build a knowledge base. Then, it parses the source code to represent it as a database entity. Additionally, Apache Lucene and fingerprint are used to support keyword-base search and structural representation of the repository, respectively. Finally, the ranker retrieves the most relevant results.

\SO has been exploited to enrich code queries, with the aim of getting relevant source code. In particular, \FaC~\cite{kim_f_2018} is a code-to-code search engine that recommends relevant  GitHub snippets to a project being developed. It is based on an alternate query technique to augment the possible retrieved results. The initial query is built from StackOverflow posts, and the additional query is performed directly on GitHub local repositories to deliver final recommendation items.

Differently from the work that proposes tools being able to provide developers with specific recommendations, Korotaev~\etal~\cite{korotaev_method_2018} introduce a GRU-based recurrent neural network (RNN) to build a universal recommender system. To this end, the approach supports the recommendation phase using a client-server architecture equipped with different components. The data collection and processing phases are conducted, taking into consideration user's behavior. The data mining module is used to feed a GRU-based RNN. To support user profiling, the approach uses ontologies by building an external knowledge representation module. The proposed network outperforms the long-short term memory (LSTM) technique concerning accuracy.

\revised{\MM has been built following a series of recommender systems developed through the CROSSMINER project~\cite{di_rocco_development_2021}. Rather than recommending source code, or libraries, \MM provides modelers with artifacts related to metamodeling activities, using a collaborative filtering technique. Such a technique has been successfully exploited to build recommdender systems to suggest API calls~\cite{9359479} and third-party libraries~\cite{NGUYEN2019110460}. \MM processes input data with four different encoding techniques. Its internal design is also tailored to compute similarity among metamodels in an efficient way.}

\subsection{Application of ML in MDE} \label{sec:MLinMDE}

In recent years, we have witnessed a proliferation of Artificial Intelligence (AI) in various aspects of human life. In the MDE domain, though learning algorithms have been successfully applied to tackle various issues, 
the adoption of AI and Machine Learning (ML) techniques in this domain is still in its infancy. This section recalls some of the most important work in this topic.

\revised{Mussbacher \etal \cite{10.1145/3417990.3421396} conducted an initial investigation on Intelligent Modeling Assistants (IMAs) using a comprehensive assessment grid. To elicit critical IMAs features, the authors analyzed the well-founded Reference Framework for Intelligent Modeling Assistance (RF-IMA). The main finding of the work is that existing IMAs obtain low scores in the extracted features and they can be further enhanced in terms of performance as well as in the underpinning structure.}
	
Breuker~\cite{6758697} reviews the main modeling languages used in ML as well as inference algorithms and corresponding software implementations. The aim of this work is to explore the opportunities of defining a DSML for probabilistic modeling.
To allow developers to design solutions that solve machine learning-based problems by automatically generating code for other technologies as a transparent bridge, a language and a tech-nology-independent development environment are introduced in~\cite{garcia2015towards}. A similar tool named OptiML has been proposed~\cite{sujeeth2011optiml}, aiming to bridge the gap between ML algorithms and heterogeneous hardware to provide a productive programming environment.

In collaborative modeling, Barriga \etal~\cite{DBLP:conf/models/BarrigaRH18} propose the adoption of an unsupervised and Reinforcement Learning (RL) approach to repair broken models, which have been corrupted because of conflicting changes. The main intent is to potentially reach model repairing with human-quality without requiring supervision. 

\AU~\cite{8906979} can be considered as the first attempt to classify metamodels exploiting a Machine Learning algorithm. The approach is built of top of a 
neural network to learn from labeled metamodels and classify unlabeled data. Despite its simplicity, the tool is efficient and on a 
small dataset, it classifies the metamodels, obtaining high prediction accuracy. In a recent work~\cite{NGUYENmemoCNN2020}, we further improved the performance by employing a convolutional neural network to classify metamodels.


Heterogeneity issues in customizable recommender systems have been analyzed by involving two different use cases \cite{solomon_heterogeneity_2016}. The participants had to tune the system according to their preferences. In the first session, users configured a travel recommender by means of different facets of the trip, \ie costs, food, and location. The second use case involved a personal exercise recommender system for the training activity. The results show that even homogeneous groups of users select different system configurations. Thus, a tailored recommender system might consider the mental model of the target users, namely their preferences and custom algorithms.

Blended recommending \cite{loepp_blended_2015} introduces a similar strategy embedded in a movie recommender.  It implements several filtering techniques used in the domain, \ie content-based filtering and collaborative filtering. Using the blended recommending strategy, users can specify a recommendation algorithm as well as refine its parameters in a hybrid filtering fashion. In this landscape, our work aims to cope with these challenges by promoting the adoption of a low-code platform. To our best knowledge, this is the first attempt to use such a technology in this domain.
	 
A recent work~\cite{stephan2019towards} envisioned a new idea for supporting the modeler with step-wise guidance or entire model examples. The proposed approach involves model clone detection techniques to find similar metamodels to the one that the modeler is defining. In contrast to our approach, the discovery of similar metamodels does not include language syntaxes, but it relies on the Simone model clone detector.


\vspace{-.4cm}
\section{Conclusions and future work}
\label{sec:Conclusions}
In this paper, we introduced \MM, a novel approach that uses a context-aware collaborative filtering technique to support the modeler in completing the specification of a metamodel.
By encoding metamodels and their contents in four different schemes, we built rating matrices and applied a syntactic-based similarity function to predict missing items, \ie classes and structural features.
An evaluation on two independent datasets, \ie \textbf{D$_1$} and \textbf{D$_2$}, and four encoding schemes, \ie SE$_{s}$, IE$_{s}$, SE$_{c}$, and IE$_{c}$, exploiting ten-fold cross-validation demonstrates that the tool is able to provide decent recommendations. 

We plan to extend \MM by adding other similarity functions, \eg structural and semantic based methods. Moreover, we can improve the encoding sche\-mes by introducing Natural Language Pre-processing (NLP) techniques. We will augment additional information to the recommendation outcomes, \eg type, cardinality.
\revised{Afterward, we are going to conduct a proper user study with the involvement of modelers to evaluate the usability of \MM. Last but not least, now that we have validated the algorithmic accuracy of the proposed technique, we will integrate the conceived tool into the Eclipse IDE, providing modelers with supports embedded in their development environment.}
	 


\begin{acknowledgements}
	\revised{The research described in this paper has been partially supported by the AIDOaRT Project, which has received funding from 
	the European Union's H2020-ECSEL-2020, Federal Ministry of Education, Science and Research, Grant Agreement n$^{\circ}$  101007350. We thank the anonymous reviewers for their valuable comments and suggestions that helped us improve the paper.}
\end{acknowledgements}



\balance
\bibliographystyle{abbrv}
\bibliography{main} 


\end{document}